\documentclass[11pt,english,onecolumn ]{IEEEtran}

\usepackage[T1]{fontenc}
\usepackage[latin1]{inputenc}
\usepackage{geometry}
\usepackage{amsmath,amssymb,amsthm,mathrsfs,amsfonts,dsfont}
\usepackage{graphicx}
\usepackage{epstopdf}
\usepackage{cite}
\usepackage{color}
\usepackage{mathtools}
\usepackage{cases}
\usepackage{stackengine}
\usepackage{accents}
\usepackage{caption2}
\usepackage{float}
\usepackage{verbatim}
\usepackage{subfigure}
\usepackage{multirow}
\usepackage{cancel}
\usepackage[shortlabels]{enumitem}
\usepackage{algorithm}
\usepackage[noend]{algpseudocode}

\makeatletter
\def\BState{\State\hskip-\ALG@thistlm}
\makeatother

\newtheorem{lem}{Lemma}
\newtheorem{prop}{Proposition}

\newtheorem{rem}{Remark}

\renewcommand\appendix{\par
\setcounter{section}{0}
\setcounter{subsection}{0}
\setcounter{figure}{0}
\setcounter{table}{0}
\renewcommand\thesection{ \Alph{section}}
\renewcommand\thefigure{\Alph{section}\arabic{figure}}
\renewcommand\thetable{\Alph{section}\arabic{table}}
}

\usepackage{babel}

\allowdisplaybreaks

\begin{document}

\title{SWIPT Signalling over Frequency-Selective Channels with a Nonlinear Energy Harvester: Non-Zero Mean and Asymmetric Inputs}

\author{Morteza Varasteh$^\dagger$, Borzoo Rassouli$^*$ and Bruno Clerckx$^\dagger$\\
$^\dagger$ Department of Electrical and Electronic Engineering, Imperial College London, UK.\\
$^*$ School of Computer Science and Electronic Engineering, University of Essex, UK.\\
\{m.varasteh12; b.clerckx\}@imperial.ac.uk, b.rassouli@essex.ac.uk.
\thanks{M. Varasteh and B. Clerckx are with the EEE department at Imperial College London, London SW7 2AZ, UK (email: \{m.varasteh12; b.clerckx\}@imperial.ac.uk). B. Rassouli is with School of Computer Science and Electronic Engineering, University of Essex, UK (email: b.rassouli@essex.ac.uk). This work has been partially supported by the EPSRC of the UK under grant EP/P003885/1.
}}

\maketitle

\begin{abstract}
Simultaneous Wireless Information and Power Transfer (SWIPT) over a point-to-point frequency-selective Additive White Gaussian Noise (AWGN) channel is studied. Considering an approximation of the nonlinearity of the harvester, a general form of delivered power in terms of system baseband parameters is derived, which demonstrates the dependency of the delivered power on higher order moment of the baseband channel input distribution. The optimization problem of maximizing Rate-Power (RP) region is studied. Assuming that the Channel State Information (CSI) is available at both the receiver and the transmitter, and constraining to non-zero mean Gaussian input distributions, an optimization algorithm for power allocation among different subchannels is studied. As a special case, optimality conditions for zero mean Gaussian inputs are derived. Results obtained from numerical optimization demonstrate the superiority of non-zero mean Gaussian inputs (with asymmetric power allocation in each complex subchannel) in yielding a larger RP region compared to their zero mean and non-zero mean (with symmetric power allocation in each complex subchannel) counterparts. This severely contrasts with SWIPT design under linear energy harvesting, for which circularly symmetric Gaussian inputs are optimal.
\end{abstract}

\section{Introduction}\label{Sec_Intro}
Radio-Frequency (RF) waves can be utilized for transmission of both information and power simultaneously. RF transmissions of these quantities have traditionally been treated separately. Currently, the community is experiencing a paradigm shift in wireless network design, namely unifying transmission of information and power, so as to make the best use of the RF spectrum and radiation, as well as the network infrastructure for the dual purpose of communicating and energizing \cite{Clerckx_Zhang_Schober_Wing_Kim_Vincent_arxiv}. This has led to a growing attention in the emerging area of Simultaneous Wireless Information and Power Transfer (SWIPT). As one of the primary works in the information theory literature, Varshney studied SWIPT in \cite{Varshney_2008}, in which he characterized the capacity-power function for a point-to-point discrete memoryless channel. Recent results in the literature have also revealed that in many scenarios, there is a tradeoff between information rate and delivered power. Just to name a few, frequency-selective channel \cite{Grover_Sahai_2010}, MIMO broadcasting \cite{Zhang_Keong_2013}, interference channel \cite{Park_Clerckx_2013}.

The main challenege in Wireless Power Transfer (WPT) is to increase the Direct-Current (DC) power at the output of the harvester without increasing transmit power. The harvester, known as rectenna, is composed of an antenna followed by a rectifier.\footnote{In the literature, the rectifier is usually considered as a nonlinear device (usually a diode) followed by a low-pass filter. The diode is the main source of nonlinearity induced in the system.} In \cite{Trotter_Griffin_Durgin_2009,Clerckx_Bayguzina_2016}, it is shown that the RF-to-DC conversion efficiency is a function of the rectenna's structure, as well as its input waveform (power and shape). Accordingly, in order to maximize the rectenna's output power, a systematic waveform design is crucial to make the best use of an available RF spectrum \cite{Clerckx_Bayguzina_2016}. In \cite{Clerckx_Bayguzina_2016}, an analytical model for the rectenna's output is introduced via the Taylor expansion of the diode characteristic function and a systematic design for multisine waveform is derived. The nonlinear model and the design of the waveform was validated using circuit simulations in \cite{Clerckx_Bayguzina_2016, Clerckx_Bayguzina_2017} and recently confirmed through prototyping and experimentation in \cite{Kim_Clerckx_Mitcheson}. Those works also confirm the inaccuracy of linear dependence of the rectifier's output power on its input power\footnote{The linear model has for consequence that the RF-to-DC conversion efficiency of the energy harvester (EH) is constant and independent of the harvester's input waveform (power and shape) \cite{Zhang_Keong_2013,Zeng_Clerckx_Zhang2017}.}. As one of the main conclusions, it is shown that the rectifier's nonlinearity is beneficial to the system performance and has a significant impact on the design of signals and systems involving wireless power.

The SWIPT literature has so far, to a great extent, ignored the nonlinearity of the EH and has focused on the linear model of the rectifier, e.g., \cite{Grover_Sahai_2010,Zhang_Keong_2013,Park_Clerckx_2013}. However, it is recognized that considering the harvester nonlinearity changes the design of SWIPT at the physical layer and medium access control layer \cite{Clerckx_Zhang_Schober_Wing_Kim_Vincent_arxiv}. Nonlinearity leads to various energy harvester models \cite{Clerckx_Bayguzina_2016,Boshkovska_Ng_Zlatanov_2017,Alevizos_Bletsas_2018}, new designs of modulation and input distribution \cite{Varasteh_Rassouli_Clerckx_ITW_2017,Varasteh_Rassouli_Clerckx_arxiv,Bayguzina_Clerckx_2018}, waveform \cite{Clerckx_2016}, RF spectrum use \cite{Clerckx_2016}, transmitter and
receiver architecture \cite{Clerckx_2016,Varasteh_Rassouli_Clerckx_arxiv,Kang_Kim_Kim_2018} and resource allocation \cite{Boshkovska_Ng_Zlatanov_2017,Xiong_Wang_2017,Xu_Ozcelikkale_McKelvey_2017}.
Of particular interest is the role played by nonlinearity on SWIPT signalling in single-carrier and multi-carrier transmissions \cite{Clerckx_2016,Varasteh_Rassouli_Clerckx_ITW_2017,Varasteh_Rassouli_Clerckx_arxiv,Clerckx_Zhang_Schober_Wing_Kim_Vincent_arxiv,Morsi_Jamali}. In multi-carrier transmissions, it is shown in \cite{Clerckx_2016} that inputs modulated according to the Circular Symmetric Complex Gaussian (CSCG) distributions, improve the delivered power compared to an unmodulated continuous waves. Furthermore, in \cite{Varasteh_Rassouli_Clerckx_ITW_2017}, it is shown that for an AWGN channel with complex Gaussian inputs under average power and delivered power constraints, depending on the receiver demand on information and power, the power allocation between real and imaginary components is asymmetric. As an extreme point, when the receiver merely demands for power requirements, all the transmitter power budget is allocated to either real or imaginary components. In \cite{Varasteh_Rassouli_Clerckx_arxiv,Morsi_Jamali}, it is shown that the capacity achieving input distribution of an AWGN channel under average, peak and delivered power constraints is discrete in amplitude with a finite number of mass-points and with a uniformly distributed independent phase. In multi-carrier transmission, however, it is shown in \cite{Clerckx_2016} that non-zero mean Gaussian input distributions lead to an enlarged Rate-Power (RP) region compared to CSCG input distributions. This highlights that the choice of a suitable input distribution (and therefore modulation and waveform) for SWIPT is affected by the EH nonlinearity and motivates the study of the capacity of AWGN channels under nonlinear power constraints.

Our interests in this paper lie in the apparent difference in input distribution for single-carrier and multi-carrier transmission, that is single-carrier favors asymmetric inputs \cite{Varasteh_Rassouli_Clerckx_ITW_2017}, while multi-carrier favors non-zero mean inputs \cite{Clerckx_2016}. We aim at tackling the design of input distribution for SWIPT under nonlinear constraints using a unified framework based on non-zero mean and asymmetric distributions. To that end, we study SWIPT in a multi-carrier setting subject to nonlinearities of the EH. We consider a frequency-selective channel subject to transmit average power and receiver delivered power constraints. We mainly focus on complex Gaussian inputs, where inputs of each real subchannel are independent of each other and on each real subchannel the inputs are independent and identically distributed (iid).

We are aiming at reconciling the two main observations of the previous paragraph: that is, outperforming of asymmetric Gaussian inputs and non-zero mean Gaussian inputs compared to CSCG inputs in single-carrier transmission \cite{Varasteh_Rassouli_Clerckx_ITW_2017} and multi-carrier transmission \cite{Clerckx_2016}, respectively. The contributions of this paper are listed below.
\begin{itemize}
\item First, taking the advantage of the small-signal approximation for rectenna's nonlinear output introduced in \cite{Clerckx_2016}, we obtain the general form of the delivered power in terms of system baseband parameters. It is shown that, first, unlike the linear model, the delivered power at the receiver is dependent on higher moments of the channel input, such as the first, second and forth moments. Second, the amount of delivered power on each subchannel is dependent on its adjacent subchannels.

\item Assuming non-zero mean Gaussian inputs, an optimization algorithm is introduced. Numerical optimizations reveal that for the scenarios where the receiver is interested in both information and power, simultaneously, the inputs are with non-zero mean and non-zero variance. Two important observations are made: first, that allowing the input to be non-zero mean improves the rate-power region, significantly, and second, that for receiver demands, which concerns information and power, the power allocation between real and imaginary components of each complex subchannel is asymmetric in general. These results, can be thought of as generalization of the results in \cite{Varasteh_Rassouli_Clerckx_ITW_2017} and \cite{Clerckx_2016}, where asymmetric power allocation (in flat fading channels) and non-zero mean inputs (in frequency-selective channels) are proposed, respectively, in order to achieve larger RP regions.

\item As a special scenario, we consider the optimized zero mean Gaussian inputs under the assumption of nonlinear EH. For this case, optimality conditions are derived. It is shown that (similar to non-zero mean inputs) under nonlinear assumption for the EH, the power allocation on each subchannel is dependent on other subchannels as well. Forcing the optimality conditions to be satisfied (numerically), it is observed that a larger RP region is obtained in contrast to the optimal zero mean inputs under the linear assumption for the EH.
\end{itemize}

\textit{Organization}: In Section \ref{Sec:sys_model}, we introduce the system model. In Section \ref{Sec:Prob}, the studied problem is introduced. In Section \ref{Sec:Power}, The delivered power at the output of the EH is obtained in terms of system baseband parameters. In Section \ref{Sec:RP}, the rate-power maximization over frequency-selective channels with non-zero mean Gaussian inputs is considered. As a special case, the optimality conditions for power allocation on different subchannels are obtained for zero mean Gaussian inputs. In Section \ref{Sec_Numerical}, WPT and SWIPT optimization for the studied problem is introduced and numerical results are presented. We conclude the paper in Section \ref{Sec:Conc} and the proofs for some of the results are provided in the Appendices at the end of the paper.

\textit{Notation}: Throughout this paper, random variables and their realizations are represented by capital and small letters, respectively. $\mathbb{E}[Y(t)]$ and $\mathcal{E}[Y(t)]$ denote the expectation over statistical randomness and the average over time of the process $Y(t)$, respectively, i.e.,
\begin{align}
\mathbb{E}[Y(t)]&=\int_{\infty}^{\infty} y(t)dF_{Y(t)}(y),\\
\mathcal{E}[Y(t)]&=\lim_{T\rightarrow \infty}\frac{1}{T}\int_{-T/2}^{T/2} Y(t)dt,
\end{align}
where $F_{Y(t)}(y)$ denotes the Cumulative Distribution Function (CDF) of the process $Y(t)$. $\otimes$ denotes circular convolution. The standard CSCG distribution is denoted by $\mathcal{CN}(0,1)$. Complex conjugate of a complex number $c$ is denoted by $c^{*}$. $\Re\{\cdot\}$ and $\Im\{\cdot\}$ are real and imaginary operators, respectively. For a complex random variable $V$, we denote $\mathbb{E}[|V|^4]=Q$, $\mathbb{E}[|V|^2]=P$, $\mathbb{E}[V^2]=\bar{P}$, $\mathbb{E}[V]=\mu$ and $\mathbb{E}[|V-\mu|^2]=\sigma^2$. The moments corresponding to real and imaginary components of $V$ are represented by subscripts $r$ and $i$, respectively, i.e., $\mathbb{E}[\Re\{V\}^4]=Q_{r}$, $\mathbb{E}[\Re\{V\}^2]=P_{r}$, $\mathbb{E}[\Re\{V\}]=\mu_{r}$ and $\mathbb{E}[|\Re\{V\}-\mu_r|^2]=\sigma_r^2$ and similarly for imaginary counterparts. $(\cdot)_N$ denotes remainder of the argument with respect to $N$. $\delta_k=1$ for $k=0$ and zero elsewhere. $\mathrm{sinc}(t)=\frac{\sin(\pi t)}{\pi t}$ and $\delta^{l}_{k}\triangleq 1-\delta_{l-k}$. $f^x$ denotes the partial derivative of the function $f$ with respect to $x$, i.e., $\frac{\partial f}{\partial x}$. The vector $[V_0,\ldots,V_{N-1}]$ is represented by $\pmb{V}^N$. Throughout the paper, complex subchannels and their real/imaginary components are referred to as c-subchannels and r-subchannels, respectively.

\section{System Model}\label{Sec:sys_model}
Considering a point-to-point $L$-tap frequency-selective AWGN channel, in the following, we explain the operation of the transmitter and the receiver.

\subsection{Transmitter}
The transmitter utilizes Orthogonal Frequency Division Multiplexing (OFDM) to transmit information and power over the channel. Let $\pmb{V}^N$ denote the modulated Information-Power (IP) complex symbols over $N$ sub-carriers (c-subchannels), occupying the overall bandwidth of $f_w$ Hz and being uniformly separated by $f_w/N$ Hz. Inverse Discrete Furrier Transform (IDFT)\footnote{In this paper we consider $X_k=\frac{1}{\sqrt{N}}\sum_{n=0}^{N-1}x[n]e^{-j\frac{2\pi nk}{N}}$ and $x[n]=\frac{1}{\sqrt{N}}\sum_{k=0}^{N-1}X_ke^{j\frac{2\pi nk}{N}}$ for DFT and IDFT definitions, respectively.} is applied over IP symbols $\pmb{V}^N$ and Cyclic Prefix (CP) is added to produce the time domain signal $X[n]$ given by
\begin{align}
 X[n+L] &= \frac{1}{\sqrt{N}}\sum_{k=0}^{N-1}V_ke^{\frac{j2\pi nk}{N}}, ~n=0,...,N-1.
\end{align}
Next, the signal
\begin{align}\label{eqn_1}
 X(t)=\sum_{n=0}^{N+L-1}X[n]\text{sinc}(f_wt-n),
\end{align}
is upconverted to the carrier frequency $f_c$ and is transmitted over the channel.

\subsection{Receiver}
The filtered received RF waveform at the receiver is modelled as
\begin{align}
  Y_{\text{rf}}(t) &=\sqrt{2}\Re\left\{Y(t)e^{j2\pi f_ct}\right\},
\end{align}
where $Y(t)$ is the baseband equivalent of the channel output with bandwidth $[-f_w/2,f_w/2]$ Hz. In order to guarantee narrowband transmission, we assume that $f_c\gg2f_w$.

\textit{Delivered Power}: The power of the signal $Y_{\text{rf}}(t)$ (denoted by $P_{\text{dc}}$) is harvested using a rectenna. The delivered power is modelled as
\begin{align}\label{eqn_2}
P_{\text{dc}}=\mathbb{E}\mathcal{E}[k_2Y_{\text{rf}}(t)^2 + k_4 Y_{\text{rf}}(t)^4],
\end{align}
where $k_2$ and $k_4$ are constants\footnote{The reader is referred to \cite{Clerckx_Bayguzina_2016} for detailed explanations of the model. Also note that according to \cite{Clerckx_2016}, rectenna's output is in the form of current with unit Ampere. However, since power is proportional to current, with abuse of notation, we refer to the term in (\ref{eqn_2}) as power.}.

\textit{Information Receiver}: The signal $Y_{\text{rf}}(t)$ is downconverted and sampled with sampling frequency $f_w$ producing $Y[m]\triangleq Y(m/f_w)$ given by
\begin{align}\label{eqn_3}
  Y[m]&=\sum\limits_{d=0}^{L-1} \tilde{h}_dX[m-d]+Z[m],~m=L,\ldots, N+L-1,
\end{align}
where $Z[m]$ represents a sample of the additive noise at time $t=m/f_w$. $\tilde{h}_d$ is the $d^{\text{th}}$ c-subchannel tap and $X[m-d]$ is a sample of the signal $X(t)$ given in (\ref{eqn_1}) at time $(m-d)/f_w$.

Considering one OFDM block, the receiver discards the CP and converts the $N$ symbols back to the frequency domain by applying DFT on (\ref{eqn_3}), such that
\begin{align}\label{eqn_4}
  Y_{l}&= h_lV_l+W_l,~l=0,\cdots,N-1,
\end{align}
where $Y_{l},l=0,\cdots,N-1$ is the DFT of $Y[m],m=L,...,L+N-1$. $h_l,~V_l$ and $W_l$ are DFTs of the extended channel vector $\tilde{\mathbf{h}}\triangleq [\tilde{h}_0,\cdots,\tilde{h}_{L-1},0,\cdots,0]_{1\times N}$, symbols $X[m],m=L,...,L+N-1$ (equivalently, samples of $X(t)$ at times $m/f_w$) and noise samples $Z[m],m=L,...,L+N-1$, respectively. That is,
\begin{subequations}
\begin{align}
h_l&=\frac{1}{\sqrt{N}}\sum_{n=0}^{N-1}\tilde{\mathbf{h}}\left[n\right]e^{-\frac{j2\pi nl}{N}}, ~~l=0,\cdots,N-1,\\
V_l &= \frac{1}{\sqrt{N}}\sum_{n=L}^{L+N-1}X\left[n\right]e^{-\frac{j2\pi nl}{N}}, ~~l=0,\cdots,N-1,
\end{align}
\end{subequations}
and similarly for $W_l,~l=0,\cdots,N-1$. We assume $W_l,~l=0,\cdots,N-1$ as iid and CSCG random variables with variance $\sigma_w^2$, i.e., $W_l\sim \mathcal{CN}(0,\sigma_w^2)$ for $l=0,\cdots,N-1$. The channel frequency response is assumed to be known at the transmitter.

\section{Problem statement}\label{Sec:Prob}
We aim at maximizing the rate of transmitted information, as well as the amount of delivered power at the receiver, given that the input in each c-subchannel $l=0,\ldots,N-1$ is distributed according to a non-zero mean complex Gaussian distribution. We also assume that in each c-subchannel the real and imaginary components are independent. Accordingly, the optimization problem consistes in the maximization of the mutual information between the channel input $\pmb{V}^N$ and the channel output $\pmb{Y}^N$ (see eq. \ref{eqn_4}) under an average power constraint at the transmitter and a delivered power constraint at the receiver, such that, $V_{lr}\sim \mathcal{N}(\mu_{lr},P_{lr}-\mu_{lr}^2)$ and $V_{li}\sim \mathcal{N}(\mu_{li},P_{li}-\mu_{li}^2)$ for $l=0,\cdots,N-1$. Hence, we have
\begin{equation}\label{eqn_21}
\begin{aligned}
& \underset{ \mu_{lr},\mu_{li},P_{lr},P_{li},~l=0,\ldots N-1}{\text{max}}
& & I\left(\pmb{V}^N;\pmb{Y}^N\right) \\
& \text{s.t.}
& & \left\{\begin{array}{l}
      \sum_{l=0}^{N-1}P_l\leq P_a \\
      P_{\text{dc}}\geq P_d
    \end{array}\right.,
\end{aligned}
\end{equation}
where $P_l=P_{lr}+P_{li}$ and $\mu_l=\mu_{lr}+j\mu_{li}$ are the average power and mean of the $l^{\text{th}}$ c-subchannel, respectively. $P_a$ is the available power budget at the transmitter. $P_d$ is the minimum amount of average delivered power at the receiver. Maximization is taken over all the means $\mu_{lr},~\mu_{li}$ and powers $P_{lr},~P_{li}$ ($l=0,\ldots,N-1$) of independent complex Gaussian inputs $\pmb{V}^N$, such that the constraints are satisfied.

\section{Power metric in terms of channel baseband parameters}\label{Sec:Power}
In this section, we study the delivered power at the receiver based on the model in (\ref{eqn_2}). Note that most of the communication processes, such as, coding/decoding, modulation/demodulation, etc, are done at the baseband. Therefore, from a communication system design point of view, it is most preferable to have baseband equivalent representation of the system. Henceforth, in the following Proposition, we derive the delivered power $P_{\text{dc}}$ at the receiver in terms of system baseband parameters. For brevity of representation, we neglect the delivered power from CP, and also we assume that $N$ is odd (calculations can be easily extended to even values of $N$, following similar steps). The following proposition, expresses the delivered power $P_{\text{dc}}$ in (\ref{eqn_2}) in terms of the channel and its input baseband parameters.
\begin{prop}\label{Lemma1}
Given that the inputs on each r-subchannel are iid and that the inputs on different r-subchannels are independent, the delivered power $P_{\text{dc}}$ at the receiver can be expressed in terms of the channel baseband parameters and statistics of the channel input distribution as
\begin{align}\nonumber
P_{\text{dc}}&=\sum\limits_{l=0}^{N-1}\Bigg\{\alpha_lQ_l+\Big(\beta_l+g(P_l)\Big)P_l+\eta+\Re\bigg\{\bar{P}_l\sum_{k=1}^{\frac{N-1}{2}}\mu_{(l+k)_N}^*\mu_{(l-k)_N}^*\Phi_{l,k}\bigg\}\\\label{eqn_39}
&+\delta_{(N-1)}^l\cdot\sum_{k=1}^{\frac{N-1}{2}}\!\!\!\sum_{\substack{m=l+1\\m\neq(l+k)_N\\m\neq (l-k)_N}}^{N-1}\!\!\!\!\!\Re\Big\{\mu_{l}\mu_{m}\mu_{(l-k)_N}^*\mu_{(m-k)_N}^* \Psi_{l,m,k}\Big\}\Bigg\} \triangleq\sum\limits_{l=0}^{N-1}f_{ib}(Q_l,P_l,\bar{P}_l,\mu_l,h_l,N),
\end{align}
where $N$ is odd and $\alpha_l,~\beta_l,~\gamma_{l,m}$, $\eta$, $\Phi_{l,k}$, $\Psi_{l,n,k}$ and $g(P_l)$ are defined as
\begin{subequations}
\begin{align}\label{eqn_5}
\alpha_l&=\frac{3k_4}{4N}(|h_l|^4+|h_l^u|^4),\\\label{eqn_6}
 \beta_l&=k_2|h_l|^2+3k_4\sigma_w^2\left(|h_l|^2+|h_l^u|^2\right), \\\label{eqn_7}
 \gamma_{m,l}&=\frac{3k_4}{N}(|h_l|^2|h_m|^2+|h_l^u|^2|h_m^u|^2),\\
 \eta&=k_2\sigma_w^2+3Nk_4\sigma_w^4,\\
 \Phi_{l,k}&=\frac{3k_4}{2N}\big(h_l^2h_{(l+k)_N}^* h_{(l-k)_N}^*+{h_l^u}^2h_{(l+k)_N}^{u*}h_{(l-k)_N}^{u*}\big),\\
 \Psi_{l,m,k}&=\frac{3k_4}{N}\big(h_lh_m^{*}h_{(l-k)_N}^*h_{(m-k)_N}+h_l^uh_m^{u*}h_{(l-k)_N}^{u*}h_{(m-k)_N}^{u}\big),\\
 g(P_l)&=\delta^l_{N-1}\sum_{m=l+1}^{N-1}\gamma_{m,l}P_m,
\end{align}
\end{subequations}
with $h_l^u,~l=0,\cdots,N-1$ being the samples of the channel at times between two consecutive information samples (for more details see Appendix\ref{Sec:Channel_u}).
\end{prop}
\textit{Proof}: See Appendix\ref{app:2}.

\begin{rem}\label{rem_1}
We note that as also mentioned in Proposition \ref{Lemma1}, the delivered power is based on the assumption that the inputs on different r-subchannels are independent as well as being iid on each r-subchannel. Obtaining a closed form expression for the delivered power $P_{\text{dc}}$ at the receiver when the inputs on different r-subchannels are not iid is cumbersome. This is due to the fact that the fourth moment of the received signal $Y_{\text{rf}}(t)$ creates dependencies among the inputs of different r-subchannels. As another point, we note that in the calculations for the delivered power in Proposition \ref{Lemma1}, we neglect the delivered power from CP. This along with the aforementioned assumptions on the input distributions, bears the fact that the real delivered power (based on the introduced model in (\ref{eqn_2})) is larger than (\ref{eqn_39}). Indeed, the subscript $ib$ in (\ref{eqn_39}) stands for inner bound in order to express this point.
\end{rem}

\begin{rem}
Note that similar results in \cite{Varasteh_Rassouli_Clerckx_ITW_2017} are reported for single-carrier AWGN channel, where the delivered power is dependent on higher moments of the channel input.  In \cite{Clerckx_2016}, superposition of deterministic and CSCG signals are assumed for multi-carrier transmissions with the assumption that the receiver utilizes power splitter. Part of the signal is used for power transfer and the other part is used for information transmissions\footnote{We note that the model considered for signal transmission in this paper is different from the multi-subband orthogonal transmission considered in \cite{Clerckx_2016}.}. In comparison to the results in \cite{Clerckx_2016}, we note that, here, the channel input is generalized in the sense that it allows asymmetric power allocation across all r-subchannels. Also, at the receiver, no power splitter is assumed\footnote{This scenario considered in this paper can be considered as an optimistic upperbound on the system performance, since (so far) in practice, it is not possible to decode information and harvest power from the same signal, jointly.}.
\end{rem}

\section{Rate-Power Maximization Over Gaussian Inputs}\label{Sec:RP}
In this section, we consider the SWIPT optimization problem in (\ref{eqn_21}). We obtain the optimality conditions in their general form (assuming non-zero mean inputs) to be used in Section \ref{Sec_Numerical} in order to obtain (locally) optimal power allocations for different r-subchannels. In order to better understand the problem, the optimality conditions are specialized for zero mean Gaussian inputs, analytically.

\subsection{SWIPT with non-zero mean complex Gaussian inputs}
Assuming that the inputs of c-subchannels $\pmb{V}^N$ are in general with non-zero mean, the problem in (\ref{eqn_21}) can be rewritten as follows
\begin{equation}\label{eqn_23}
\begin{aligned}
& \underset{ \substack{P_{lr},P_{li},\mu_{lr},\mu_{li}\\l=0,...,N-1}}{\text{max}}
& & \!\!\sum\limits_{l=0}^{N-1}c_0\big(\log (1+a_l\sigma_{lr}^2)+\log(1+a_l\sigma_{li}^2)\big)\\
& \text{s.t.}
& & \!\!\!\!\!\!\!\!\!\!\!\!\!\!\!\!\!\left\{\begin{array}{l}
      \sum_{l=0}^{N-1}P_l\leq P_a, \\
       \sum_{l=0}^{N-1} f_{ib}(P_l,\bar{P_l},\mu_l,h_l,N)\geq P_d,\\
       \sigma_{lr}^2\geq 0 ,\sigma_{li}^2\geq 0,~l=0,...,N-1
    \end{array}\right.,
\end{aligned}
\end{equation}
where $c_0=\frac{f_w}{2N}$, $a_l=\frac{2N|h_l|^2}{f_w\sigma_w^2}$, $\sigma_{lr}^2=P_{lr}-\mu_{lr}^2$, $\sigma_{li}^2=P_{li}-\mu_{li}^2$. Note that for a Gaussian distribution in the function $f_{ib}(\cdot)$, we have $Q_l=3(P_{lr}^2+P_{li}^2)-2(\mu_{li}^4+\mu_{lr}^4)+2P_{lr}P_{li}$, $\bar{P}_l=P_{lr}-P_{li}+2j\mu_{lr}\mu_{li}$.

In Section \ref{Sec_Numerical}, we consider the numerical optimization of problem (\ref{eqn_23}) by considering its Lagrangian\footnote{The problem in (\ref{eqn_23}) is not convex and any solution obtained from solving the dual problem is in general a local optima.}. The KKT conditions for problem (\ref{eqn_23}) are detailed in Appendix\ref{app_KKT}. As it can be seen from the KKT conditions in Appendix\ref{app_KKT}, unfortunately, it is cumbersome to derive analytical results on the optimal solution of problem (\ref{eqn_23}). However, it can be shown that for the optimal solution, the average power constraint is satisfied with equality (see Appendix\ref{app_KKT} for the details).

As explained in Section \ref{Sec_Numerical}, numerical results reveal that non-zero mean asymmetric complex Gaussian inputs result in larger RP region compared to their zero mean counterparts. However, in order to better understand the problem in its general form (assuming non-zero mean), it is beneficial to look into the optimality conditions of zero mean inputs.

\subsection{SWIPT with zero mean complex Gaussian inputs}
In the following, we obtain the optimality conditions for power allocation among different r-subchannels, when the input distributions are complex Gaussian with zero mean and with independent components.

\begin{lem}\label{Lem_48}
If $\pmb{\mu}^N=\pmb{0}^N$, the optimal power allocation $\pmb{P}_{r}^{N^\star},~\pmb{P}_{i}^{N^\star}$ for problem (\ref{eqn_23}) satisfies the average power and delivered power constraints with equality, i.e.,
\begin{subequations}\label{eqn_50}
\begin{align}
  \sum_{l=0}^{N-1}P_l^{\star}&= P_a,\\\label{eqn_47}
  \sum_{l=0}^{N-1} f_{ib}(P^{\star}_l,\bar{P^{\star}_l},0,h_l,N)&= P_d,
\end{align}
\end{subequations}
with $f_{ib}(P^{\star}_l,\bar{P^{\star}_l},0,h_l,N)=\alpha_lQ^{\star}_l+\Big(\beta_l+g(P^{\star}_l)\Big)P^{\star}_l+\eta$. Also for the optimal vectors $\pmb{P}_{r}^{N^\star},~\pmb{P}_{i}^{N^\star}$ we have
\begin{subequations}\label{eqn_48}
\begin{align}\label{eqn_48_1}
P_{lr}^{\star}\cdot\left(\lambda_1-G_l(\pmb{P}_{r}^{N^{\star} },\pmb{P}_{i}^{N^{\star}})\right)&=0,~l=0,...,N-1,\\\label{eqn_48_2}
P_{li}^{\star}\cdot\left(\lambda_1-G_l(\pmb{P}_{i}^{N^{\star}},\pmb{P}_{r}^{N^{\star}})\right)&=0,~l=0,...,N-1,
\end{align}
\end{subequations}
with
\begin{align}\label{eqn_51}
G_l(\pmb{P}_{r}^{N},\pmb{P}_{i}^{N})\triangleq \frac{c_1a_l}{1+a_lP_{lr}}+6\lambda_2 \alpha_lP_{lr}+\lambda_2(2\alpha_l P_{li}+\beta_l+g_1(P_{l})),
\end{align}
for some
\begin{subequations}
\begin{align}\label{eqn_53}
  \lambda_1&\geq \max_{l=0,\ldots,N-1}\{G_l(\pmb{P}_{r}^{N^{\star} },\pmb{P}_{i}^{N^{\star}}),G_l(\pmb{P}_{i}^{N^{\star}},\pmb{P}_{r}^{N^{\star}})\},\\
  \lambda_2&\geq0,
\end{align}
\end{subequations}
and $g_1(P_l)\triangleq\sum_{\substack{m=0\\m\neq l}}^{N-1}\gamma_{m,l} P_m$. For $\lambda_2=0$, the optimal power allocations are simplified to waterfilling solution, i.e.,
\begin{align}
P_{lr}^{\star}=P_{li}^{\star}=\max\left\{0,\frac{c_1}{\lambda_1}-\frac{1}{a_l}\right\},~\text{for}~l=0,\cdots,N-1.
\end{align}
\end{lem}
\textit{Proof}: See Appendix\ref{app_KKT_zeromean}.

\begin{figure}
\begin{centering}
\includegraphics[scale=0.4]{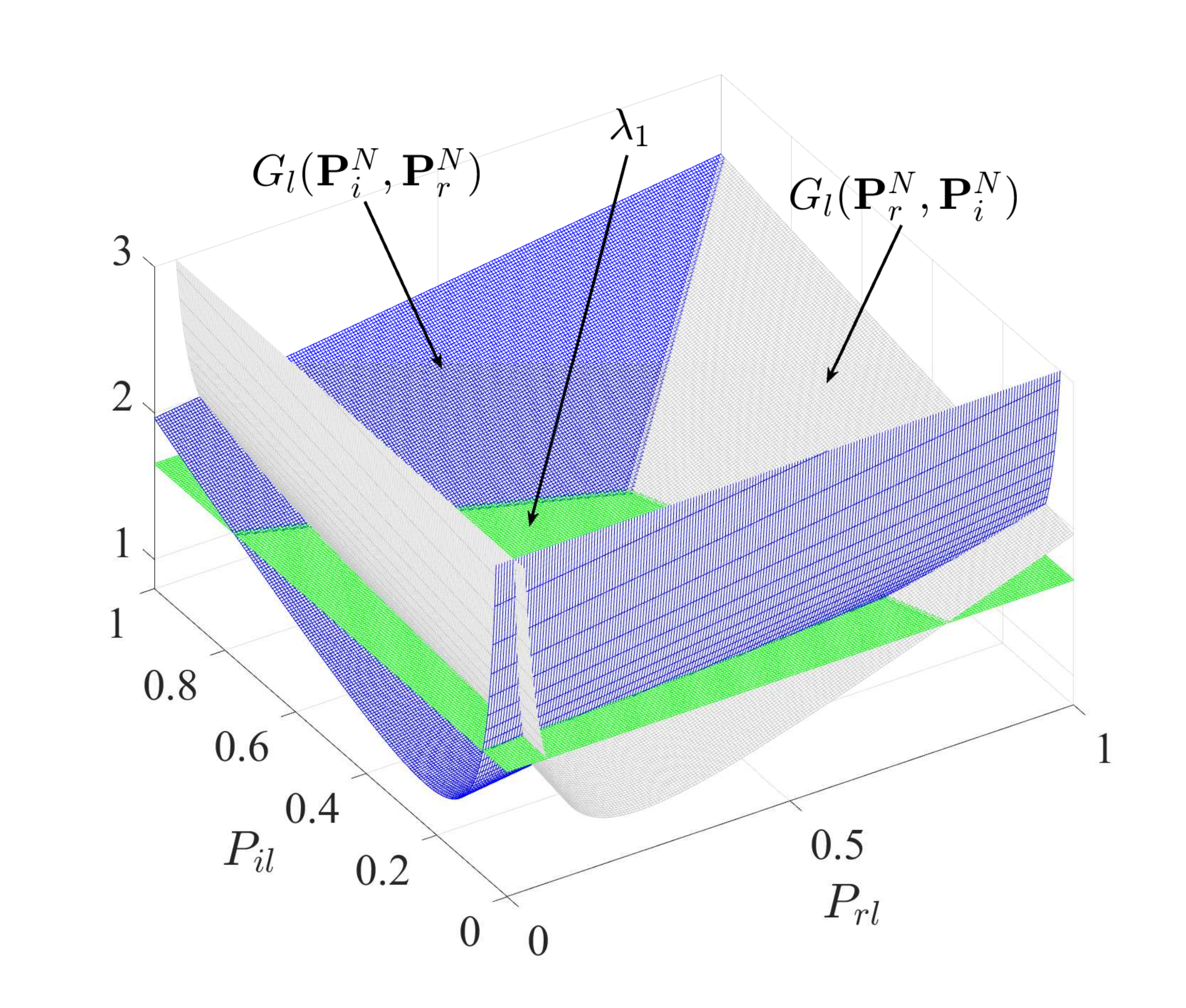}
\caption{Representation of intersection of $\lambda_1=1.6529$ with the functions $G_l(\pmb{P}_{r}^{N },\pmb{P}_{i}^{N})$ and $G_l(\pmb{P}_{i}^{N},\pmb{P}_{r}^{N})$, defined in (\ref{eqn_51}) with the parameters $c_1=0.0801,~a_l=2250,~\lambda_2=0.0498,~\alpha_l=4.9857,~\beta_l=1.6484$ and $g_1(P_l)=6.2822$. The reported parameters here correspond to the optimal solution of the strongest c-subchannel considered in Section \ref{Sec_Numerical} with average power constraint $P_a=1$, delivered power constraint $P_d=3.5716$ and noise variance $\sigma_w^2=0.1$.}\label{Fig_1}
\par\end{centering}
\vspace{0mm}
\end{figure}

\begin{rem}\label{rem_10}
Note that the delivered power in the $l^{\text{th}}$ c-subchannel for zero mean Gaussian inputs, i.e., \[f_{ib}(P^{\star}_l,\bar{P^{\star}_l},0,h_l,N)=\alpha_lQ^{\star}_l+\Big(\beta_l+g(P^{\star}_l)\Big)P^{\star}_l+\eta,\]
is dependent on other c-subchannels through $g(P^{\star}_l)$\footnote{Note that for zero mean inputs with nonlinear EH, $P_{lr}=P_{lr}^{\star},P_{li}=P_{li}^{\star}$ yields the same delivered power/ transmitted information as $P_{lr}=P_{li}^{\star},P_{li}=P_{lr}^{\star}$.}. This is in contrast with the linear model, where the delivered power is obtained as $|h_l|^2P_l+\sigma_w^2$.
\end{rem}

\begin{figure}
\begin{centering}
\includegraphics[scale=0.4]{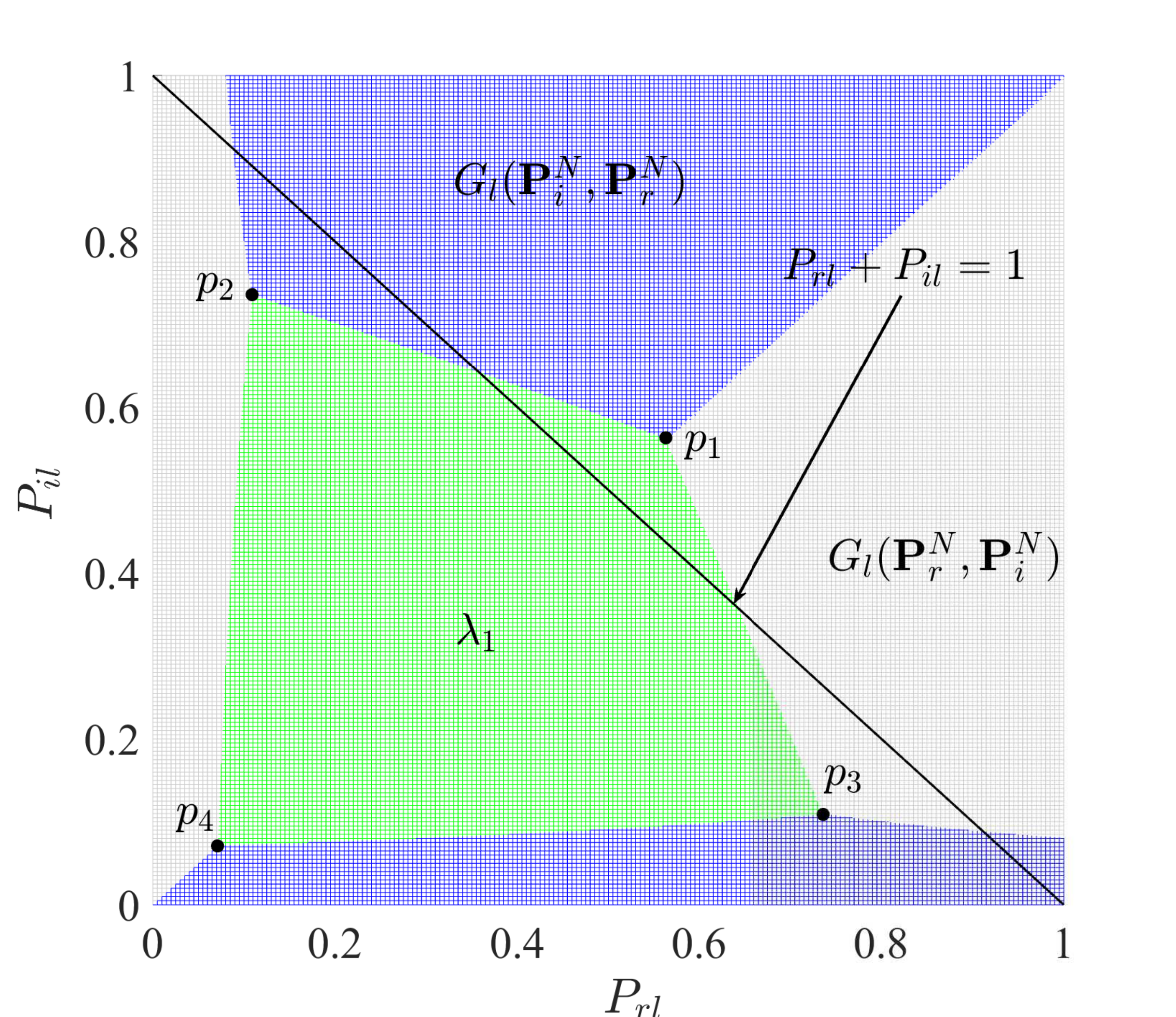}
\caption{Illustration of Figure \ref{Fig_1} from top view (along $z-$axis). There are $4$ solutions denoted by $p_1,~p_2,~p_3$ and $p_4$, where point $p_1$ is not admissible due to contradicting the average power constraint $P_a=1$.}\label{Fig_110}
\par\end{centering}
\vspace{0mm}
\end{figure}

\begin{rem}
The optimality conditions of Lemma \ref{Lem_48} in (\ref{eqn_48}) can be interpreted as follows. The functions $G_l(\pmb{P}_{r}^{N},\pmb{P}_{i}^{N}),~l=0,\ldots,N-1$ are positive and convex (the Hessian matrix is positive definite). Also note that $G_l(\pmb{P}_{i}^{N},\pmb{P}_{r}^{N})$ is a mirrored version of $G_l(\pmb{P}_{r}^{N},\pmb{P}_{i}^{N})$ with respect to the surface $P_{lr}=P_{li}$. Assume that $\lambda_2>0$ is given and that $\lambda_1$ is chosen as a large value (so that it satisfies (\ref{eqn_53})). Consider the intersection of the horizontal surface $\lambda_1$ with functions $G_l(\pmb{P}_{r}^{N},\pmb{P}_{i}^{N})$ and $G_l(\pmb{P}_{i}^{N},\pmb{P}_{r}^{N})$ for some index $l$. Depending on the value of $\lambda_1$ and shape of the functions $G_l$, different pairs of $(P_{lr},P_{li})$ satisfy simultaneously
\begin{align}\label{eqn_52}
\lambda_1=G_l(\pmb{P}_{r}^{N},\pmb{P}_{i}^{N})=G_l(\pmb{P}_{i}^{N},\pmb{P}_{r}^{N}).
\end{align}
The number of these solution pairs $(P_{lr},P_{li})$ for each index $l$ can be verified to vary from three to four. That is, if $\lambda_1>G_l(\pmb{0}^N,\pmb{0}^N)$, there are three solutions, and if $\lambda_1\leq G_l(\pmb{0}^N,\pmb{0}^N)$, there are four solutions for (\ref{eqn_52})\footnote{Note that $\lambda_1$ must satisfy the condition in (\ref{eqn_53}) as well.}. In Figure \ref{Fig_1}, an illustration of the intersection of the aforementioned three surfaces for a specific index $l$ is provided, where four pairs of solutions are recognized. In Figure \ref{Fig_110}, the same illustration is presented along the $z-$axis from the top. Points $p_1,~p_2,~p_3$ and $p_4$ denote the solution pairs that satisfy (\ref{eqn_52}). Note that depending on the average power constraint, some (or all) of the points $p_1,~p_2,~p_3$ and $p_4$ are not admissible (for example, here $p_1$ is not admissible). If there is no point satisfying the average power constraint, the power allocated to the corresponding c-subchannel is zero (in order to satisfy (\ref{eqn_48})). Otherwise, there are more than one set of power allocations $(P_{lr},~P_{li})$ that satisfy the optimality necessary conditions. Accordingly, the power allocation could be either symmetric (corresponding to either of the points $p_1,~p_4$) or asymmetric (corresponding to either of the points $p_2,~p_3$). Note that both points $p_2$ and $p_3$ contribute the same amount in the delivered power and transmitted information (as noted in Remark \ref{rem_10}). Therefore, they can be chosen interchangeably.
\end{rem}

\begin{rem}\label{rem_111}
The optimality conditions in (\ref{eqn_48}) can be solved numerically using programming for solving nonlinear equations with constraints ($P_{lr},P_{li}\geq 0$ for $l=0,\ldots,N-1$). In Section \ref{Sec_Numerical}, it is observed through numerical optimization that for mere WPT purposes (equivalently large values of $\lambda_2$) all the available power at the transmitter is allocated to either real or imaginary component of the strongest c-subchannel. Additionally, note that, (for zero mean Gaussian inputs), although optimized for WPT, the amount of transmitted information is never zero.
\end{rem}

\section{Numerical Optimization}\label{Sec_Numerical}
In this section, we provide numerical results regarding the power allocation for different r-subchannels under a fixed average power and different delivered power constraints in order to obtain different RP regions corresponding to different types of complex Gaussian inputs introduced earlier.

\subsection{Non-zero mean inputs}\label{Sec_NZM}
We note that, the optimization problem in (\ref{eqn_21}) is not convex, and accordingly, the final solution (obtained via numerical optimization) is in general a local stationary point. Due to nonconvexity of the studied problem, the final solution is dependent on the initial starting point. In order to alleviate the effect of the initial point, in our optimization, we first focus on the WPT aspect of the optimization problem with deterministic input signals\footnote{We note that although we first optimize over deterministic signals for WPT, optimizing over means and powers for SWIPT results in the same solutions, i.e., signals with almost zero variance, however, in the expense of a long simulation time. Therefore, for the starting point of the RP region, we chose the input to be deterministic.}, i.e., the variance of different r-subchannels are close to zero with a good approximation. In this case, with deterministic input signals we have $\mu_{lr}=\sqrt{P_{lr}}$, $\mu_{li}=\sqrt{P_{li}}$ for $l=0,\cdots,N-1$. Therefore, the delivered power $P_{\text{dc}}$ reads as
\begin{align}\nonumber
P_{\text{dc}}&=\sum\limits_{l=0}^{N-1}\Bigg\{\alpha_l|\mu_l|^4+\Big(\beta_l+g(|\mu_l|^2)\Big)|\mu_l|^2+\eta+\Re\bigg\{\mu_l^2\sum_{k=1}^{\frac{N-1}{2}}\mu_{(l+k)_N}^*\mu_{(l-k)_N}^*\Phi_{l,k}\bigg\}\\
&+\delta_{(N-1)}^l\cdot\sum_{k=1}^{\frac{N-1}{2}}\!\!\!\sum_{\substack{m=l+1\\m\neq(l+k)_N\\m\neq (l-k)_N}}^{N-1}\!\!\!\!\!\Re\Big\{\mu_{l}\mu_{m}\mu_{(l-k)_N}^*\mu_{(m-k)_N}^* \Psi_{l,m,k}\Big\}\Bigg\}\triangleq \sum\limits_{l=0}^{N-1}f_{\text{WPT}}(\mu_l,h_l,N).
\end{align}
Accordingly, we consider the following WPT problem
\begin{equation}\label{eqn_33}
\begin{aligned}
& \underset{ \substack{\mu_{l}\\l=0,...,N-1}}{\text{max}}
& & \!\!\sum\limits_{l=0}^{N-1}f_{\text{WPT}}(\mu_l,h_l,N)\\
& \text{s.t.}
& & \!\!\!\!\!\!\!\!\!\!\!\!\!\!\!\!\!\sum_{l=0}^{N-1}|\mu_l|^2= P_a,
\end{aligned}
\end{equation}
where the proof for the average power constraint satisfied with equality has been provided in Appendix\ref{app_KKT}. The algorithm (WPT optimization with deterministic inputs) is run for a large number of times (here we run the algorithm 1000 times) using the Matlab command \texttt{fmincon()}, and each time with a new and randomly generated initial complex mean vector $\pmb{\mu}^N$. After this stage, the solution corresponding to the highest delivered power $P_{\text{dc}}$ is chosen as the initial starting point for the SWIPT optimization.

Next, in order to solve the optimization for SWIPT, we consider the following maximization, which is the weighted summation of the transmitted information and the delivered power
\begin{equation}\label{eqn_34}
\begin{aligned}
& \underset{ \substack{P_{lr},P_{li},\mu_{lr},\mu_{li}\\l=0,...,N-1}}{\text{max}}
& & \!\!\sum\limits_{l=0}^{N-1}c_0\big\{\log (1+a_l\sigma_{lr}^2)+\log(1+a_l\sigma_{li}^2)\big\}+\lambda_2 f_{ib}(P_l,\bar{P_l},\mu_l,h_l,N)\\
& \text{s.t.}
& & \!\!\!\!\!\!\!\!\!\!\!\!\!\!\!\!\!\left\{\begin{array}{l}
      \sum_{l=0}^{N-1}P_l= P_a, \\
       P_{lr}\geq \mu_{lr}^2 ,P_{li}\geq \mu_{li}^2,~l=0,...,N-1
    \end{array}\right..
\end{aligned}
\end{equation}
We solve this problem using the Matlab command \texttt{fmincon()} as follows. $\lambda_2$ is given different values, starting from larger ones\footnote{Note that $\lambda_2$ can be interpreted as $-\frac{\partial I(V^N;Y^N)}{\partial P_{\text{dc}}}$. Therefore, intuitively a larger value of $\lambda_2$ corresponds to a higher delivered power and lower transmitted information.}. For the first round of the optimization (corresponding to the largest value of $\lambda_2$), the (locally) optimal solution obtained through previous optimization (WPT with deterministic inputs) is used as the starting point (the power for different r-subchannels is considered as $P_{lr}=\mu_{lr}^2,~P_{li}=\mu_{li}^2,~l=0,\ldots,N-1$). Similarly, for the subsequent values of $\lambda_2$, we use the solution corresponding to the previous value of $\lambda_2$. The detailed description of the optimization is presented in Algorithm \ref{euclid}.

\begin{algorithm}
\caption{SWIPT algorithm (Non-zero mean inputs)}\label{euclid}
\begin{algorithmic}[1]
\Procedure{WPT Optimization}{}
\State $M \gets$ Large number
\For {$s=1:M$}
\State Randomly initialize $\pmb{\mu}^{N}$, $\pmb{\mu}_{(s)}^{N*}=\arg \max$ (\ref{eqn_33})
\State $P_{dc,(s)}=\sum\limits_{l=0}^{N-1}f_{\text{WPT}}(\mu_{l,(s)}^{*},h_l,N)$, $S=\arg \max\limits_{s} P_{dc,(s)}$
\EndFor
\EndProcedure
\Procedure{SWIPT Optimization}{}
\State $\lambda_2\gets\lambda_{max}$, $s=1$
\State $\pmb{P}_{(s),r}^{N}\gets [\mu^{*2}_{0r,(S)},\ldots,\mu^{*2}_{(N-1)r,(S)}]$, $\pmb{P}_{(s),i}^{N}\gets [\mu^{*2}_{0i,(S)},\ldots,\mu^{*2}_{(N-1)i,(S)}]$, $\pmb{\mu}_{(s)}^{N}\gets \pmb{\mu}_{(S)}^{N*}$
\While {$\lambda_2>\lambda_{min}$}
\State $\{\pmb{P}_{(s),r}^{N*},\pmb{P}_{(s),i}^{N*}\}=\arg \max$ (\ref{eqn_34})
\State $\text{Inf}(s)\gets \sum\limits_{l=0}^{N-1}c_0\big\{\log (1+a_l\sigma_{lr,(s)}^{*2})+\log(1+a_l\sigma_{li,(s)}^{*2})\big\}$, $P_{\text{dc}}(s)\gets$ (\ref{eqn_39})
\State $s\gets (s+1)$, $\lambda_2 \gets (\lambda_2 -stp)$
\State $\pmb{P}_{(s),r}^{N}\gets \pmb{P}_{(s-1),r}^{N*}$, $\pmb{P}_{(s),i}^{N}\gets \pmb{P}_{(s-1),i}^{N*}$, $\pmb{\mu}_{(s)}^{N}\gets \pmb{\mu}_{(s-1)}^{N*}$
\EndWhile
\EndProcedure
\end{algorithmic}
\end{algorithm}

\subsection{Zero mean inputs}
In order to obtain the optimal power allocations for zero mean complex Gaussian inputs, we follow a similar approach presented in Section \ref{Sec_NZM}. The optimization problem considered here is given as
\begin{equation}\label{eqn_54}
\begin{aligned}
& \underset{ \substack{P_{lr},P_{li}\\l=0,...,N-1}}{\text{max}}
& & \!\!\sum\limits_{l=0}^{N-1}c_0\big\{\log (1+a_lP_{lr})+\log(1+a_lP_{li})\big\}+\lambda_2 f_{ib}(P_l,\bar{P_l},0,h_l,N)\\
& \text{s.t.}
& & \!\!\!\!\!\!\!\!\!\!\!\!\!\!\!\!\!\left\{\begin{array}{l}
      \sum_{l=0}^{N-1}P_l= P_a, \\
      P_{lr}\geq 0 ,P_{li}\geq 0,~l=0,...,N-1
    \end{array}\right..
\end{aligned}
\end{equation}
The optimization is explained in Algorithm \ref{euclid1}\footnote{We note that, as an alternative approach, the optimality conditions in Lemma \ref{Lem_48} can be used in order to find the optimal power allocations. To do so, solving the nonlinear equations (\ref{eqn_50}) and (\ref{eqn_48}) have to be considered with the constraints $P_{li},P_{li}\geq 0$. Accordingly, one can use the MATLAB command \texttt{fsolve()}. The optimization is initialized with a very small (in norm) power vector and each time the vector is updated until a condition on convergence is met.}.


\subsection{Numerical results}
In this section, we present the results obtained through numerical simulations. First, we focus on the optimized RP regions corresponding to different types of channel inputs. Later, we compare the constellation of optimized non-zero mean and zero mean complex Gaussian inputs on different points of their corresponding optimized RP region.

\begin{algorithm}
\caption{SWIPT algorithm (Zero mean inputs)}\label{euclid1}
\begin{algorithmic}[1]
\Procedure{SWIPT Optimization}{}
\State $\lambda_2\gets\lambda_{max}$, $M \gets$ Large number
\While {$\lambda_2>\lambda_{min}$}
\For {$t=1:M$}
\State Randomly initialize $\{\pmb{P}_{r}^{N},\pmb{P}_{i}^{N}\}$
\State $\{\pmb{P}_{(t),r}^{N*},\pmb{P}_{(t),i}^{N*}\}=\arg \max$ (\ref{eqn_54})
\State $IP(t)=$ Calculate cost in (\ref{eqn_54}) for $\{\pmb{P}_{(t),r}^{N*},\pmb{P}_{(t),i}^{N*}\}$
\EndFor
\State $S=\arg \max\limits_{t} IP(t)$, save $\{\pmb{P}_{(S),r}^{N*},\pmb{P}_{(S),i}^{N*}\}$, $\lambda_2 \gets (\lambda_2 -stp)$
\EndWhile
\EndProcedure
\end{algorithmic}
\end{algorithm}

\begin{figure}
\begin{centering}
\includegraphics[scale=0.3]{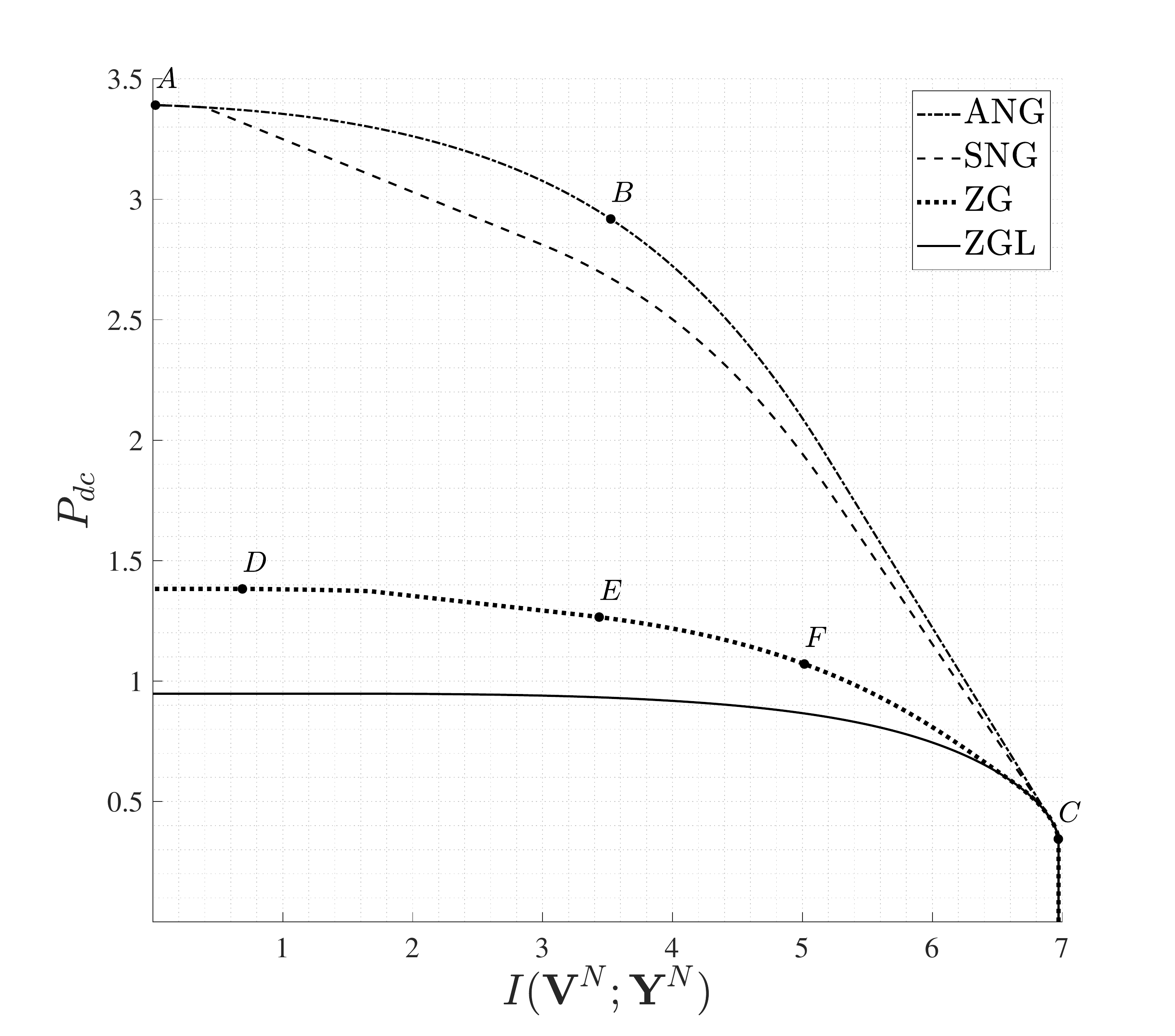}
\caption{The optimized RP regions corresponding to ANF, SNG, ZG and ZGL with average power constraint $P_a=1$ and noise variance $\sigma_w^2=0.1$.}\label{Fig_2}
\par\end{centering}
\vspace{0mm}
\end{figure}

In Figure \ref{Fig_2}, the RP regions for \textit{Asymmetric Non-zero mean Gaussian} (ANG), presented in this paper and \textit{Symmetric Non-zero mean Gaussian} (SNG) presented in \cite{Clerckx_2016} and \textit{Zero mean Gaussian} (ZG) are shown\footnote{The channel we have used for our simulations comprises $N=9$ c-subchannels with coefficients as $[ -1.2+0.1i , -0.4-1.3i , -.1-1.6i , 0.6-1.5i , -1.35-.1i , -1.1+0.2i , -0.9-.01i , 0.7+0.1i , 0.65+.01i]$.}. We also obtain the RP region corresponding to the optimal power allocations for the linear model assumption of the EH. This is done by obtaining the power allocations from \cite[Equation (9)]{Grover_Sahai_2010} for different constraints and calculating the corresponding delivered power and transmitted information. This region is denoted by \textit{Zero mean Gaussian for Linear model} (ZGL). As it is observed in Figure \ref{Fig_2}, due to the asymmetric power allocation in ANG, there is an improvement in the RP region compared to SNG. Additionally, it is observed that ANG and SNG achieve larger RP region compared to optimized ZG and that performing better than ZGL (highlighting the fact that for scenarios that the nonlinear model for EH is valid, ZGL is not optimal anymore). The main reason of improvement in the RP regions corresponding to ANG, SNG is due to the fact that allowing the mean of the channel inputs to be non-zero boosts the fourth order term (More explanations can be found in \cite{Clerckx_2016}.) in (\ref{eqn_2}), resulting in more contribution in the delivered power at the receiver.

\begin{figure}[h!]
\begin{centering}
\begin{subfigure}
\par\hfill
    \includegraphics[scale=0.213]{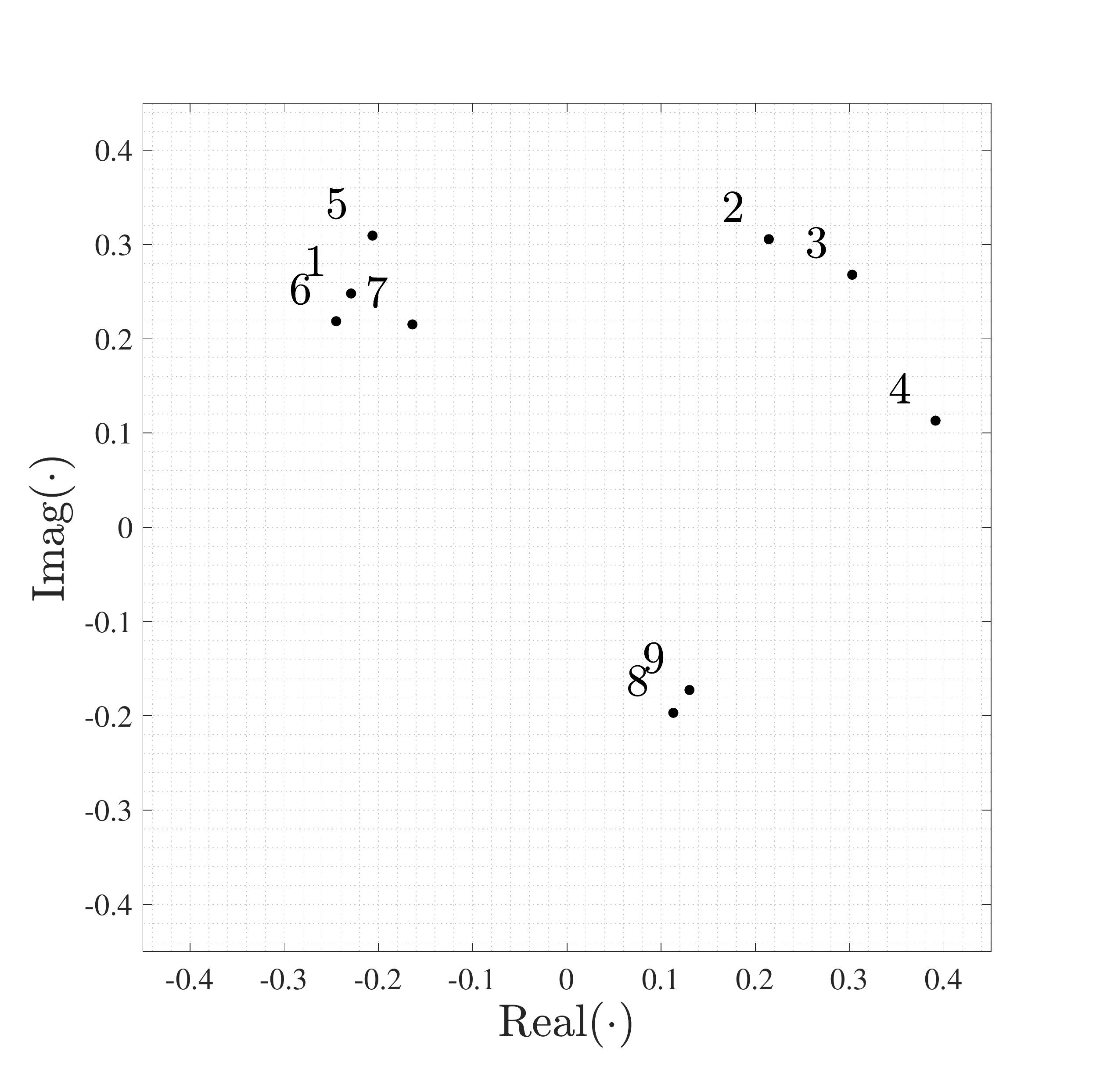}
\end{subfigure}
\begin{subfigure}
\par\hfill
  \includegraphics[scale=0.213]{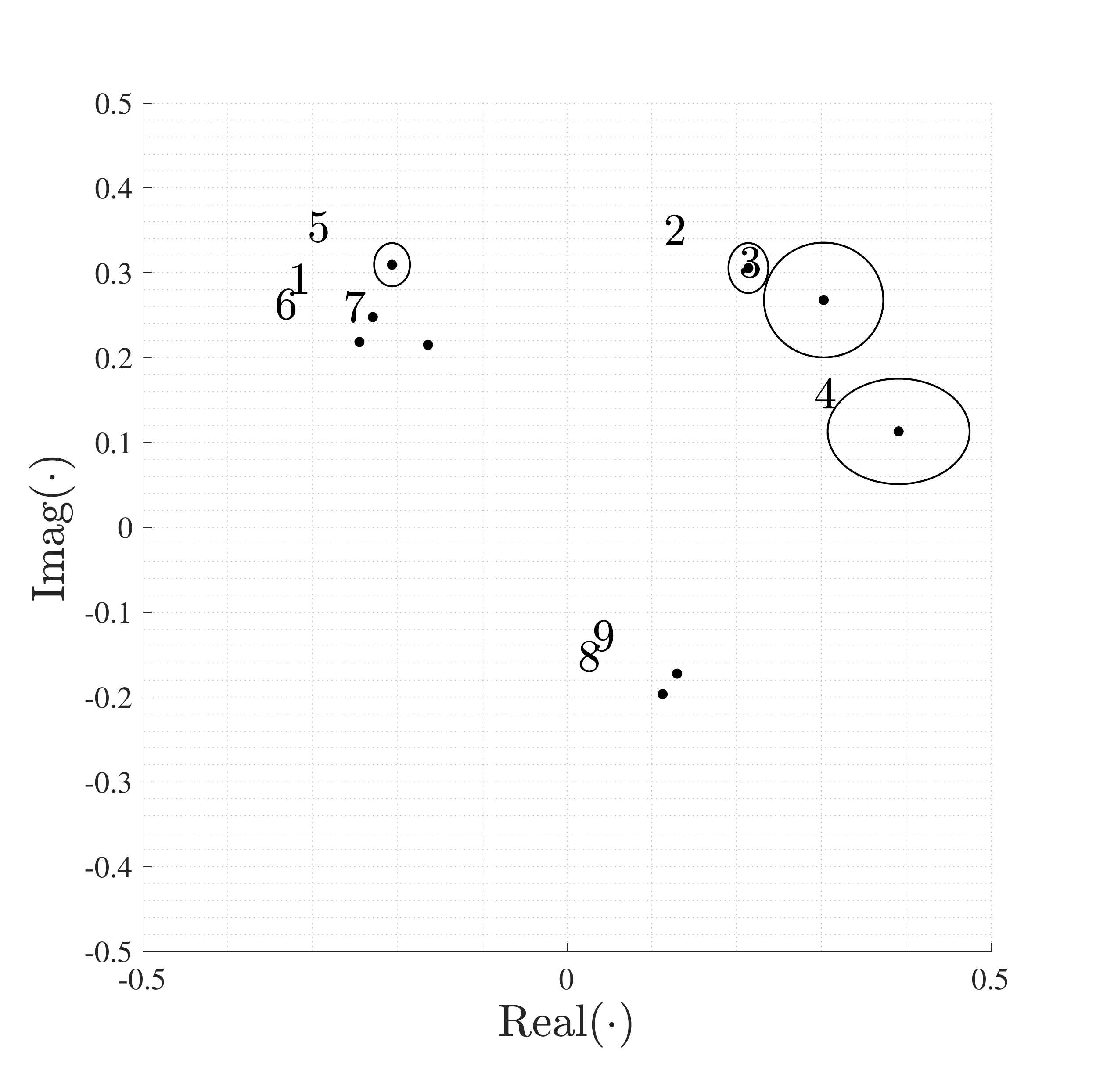}
\end{subfigure}
\begin{subfigure}
\par\hfill
  \includegraphics[scale=0.213]{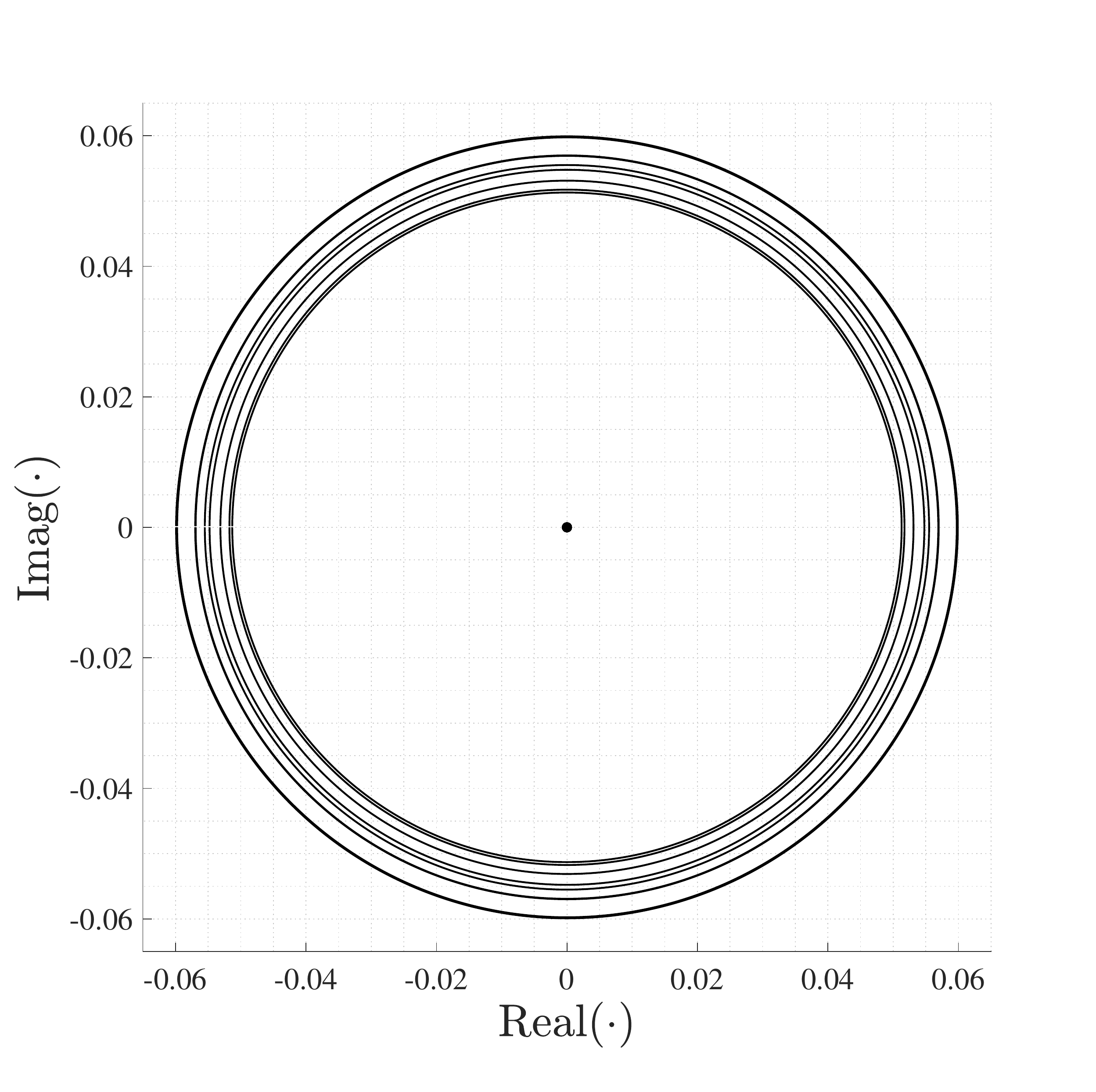}
\end{subfigure}
\caption{From left to right, the mean and variance of different r-subchannels' input corresponding to the points $A$, $B$ and $C$ in Figure \ref{Fig_2}, respectively. As we move forward from point $A$ to point $C$, the variance of different r-subchannels increase, whereas their corresponding means shrink to zero.}\label{Fig_3}
\end{centering}
\end{figure}

In Figure \ref{Fig_3}, from left to right, the optimized inputs in terms of their complex mean $\mu_l,~l=0,\ldots,8$ (represented as dots) and their corresponding r-subchannel variances $\sigma_{lr}^2,\sigma_{li}^2,~l=0,\ldots,8$ (represented as ellipses) are shown for points $A,~B$ and $C$ in Figure \ref{Fig_2}, respectively. Point $A$ represents the maximum delivered power with the zero transmitted information (note that information of a deterministic signal is zero). Point $B$ represents the performance of a typical input used for power and information transfer. Finally, point $C$ represents the performance of an input obtained via waterfilling (when the delivered power constraint is inactive). From these $3$ plots it is observed that as we move from point $A$ to point $C$, the mean of different r-subchannels decrease, however, they (means of different r-subchannels) keep their relative structure, roughly. Also, as we move to point $C$, the means of different r-subchannels get to zero with their variances increasing asymmetrically until the power allocation gets to waterfilling solution (where the power allocation between the real and imaginary components are symmetric). This result is in contrast with the results in \cite{Clerckx_2016}, where the power allocation to the real and imaginary components in each c-subchannel is symmetric. Similar results regarding the benefit of asymmetric power allocation has also been reported in \cite{Varasteh_Rassouli_Clerckx_ITW_2017} for deterministic AWGN channel with nonlinear EH.

\begin{figure}
\begin{centering}
\includegraphics[scale=0.3]{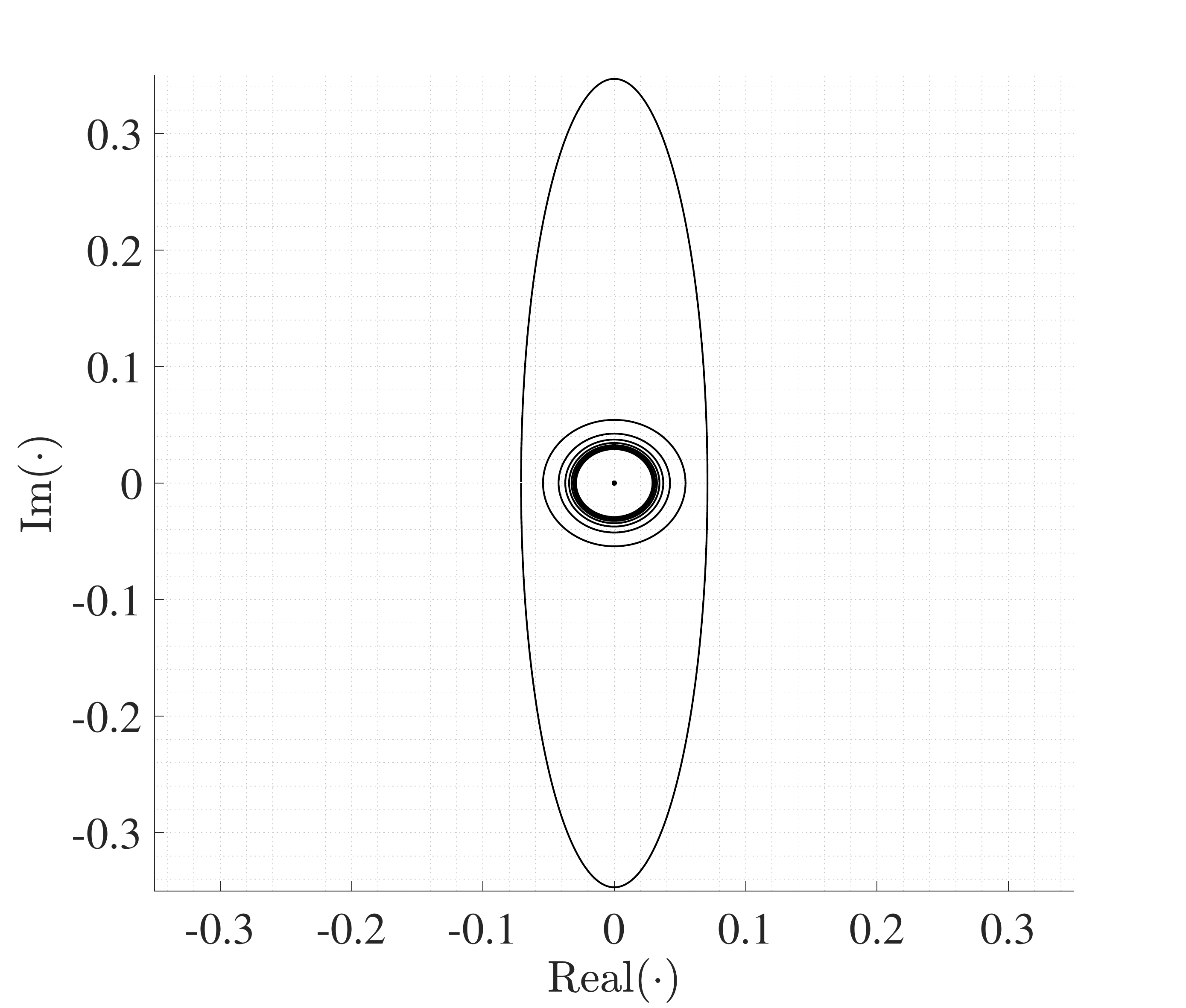}
\caption{Representation of variances of different c-subchannels corresponding to the point $E$ in Figure \ref{Fig_2}. The strongest c-subchannel receives more power compared to the other c-subchannels and the other c-subchannels attain CSCG inputs.}\label{Fig_4}
\par\end{centering}
\vspace{0mm}
\end{figure}

In Figure \ref{Fig_2}, the point $D$ corresponds to the input, where all of the c-subchannels other than the strongest one (in terms of the $\max\limits_{l=0,\ldots,N-1}|h_l|^2$) are with zero power. For the strongest c-subchannel, at point $D$, all the transmit power is allocated to either real or imaginary component of the c-subchannel. The reason for this observation is explained in Remark \ref{rem_111}. This observation is also inline with the result of \cite{Varasteh_Rassouli_Clerckx_ITW_2017}, where it is shown that for a flat fading channel, the maximum power is obtained by allocating all the transmitter power to only one r-subchannel. Note that this is different from the power allocation with the linear model (i.e. ZGL), for which  all the transmit power would also be allocated to the strongest c-subchannel to maximize delivered power but equally divided among the real and imaginary parts of the input.

In Figure \ref{Fig_4}, the variances of different r-subchannels corresponding to the point $E$ in Figure \ref{Fig_2} are illustrated. Numerical optimization reveals that, as we move from point $D$ to point $C$ (increasing the information demand at the receiver) in Figure \ref{Fig_2}, the variance of the strongest c-subchannel varies asymmetrically (in its real and imaginary components). This observation can be justified as follows. For higher values of $\lambda_2$ (equivalent to higher delivered power demands), the strongest c-subchannel receives a power allocation similar to the solutions $p_2$ or $p_3$ in Figure \ref{Fig_110}, whereas the other c-subchannels take the power allocation corresponding to the point $p_4$ in Figure \ref{Fig_110}\footnote{For very low average power constraints, it is observed that the power allocation is symmetric across all the c-subchannels. This can be justified by noting that for very low average power constraints, the admissible power allocations correspond to solutions similar to the point D in Figure \ref{Fig_110}. }. Note that the power allocation in point $C$ is the waterfilling solution. .

\begin{figure}
\begin{centering}
\includegraphics[scale=0.3]{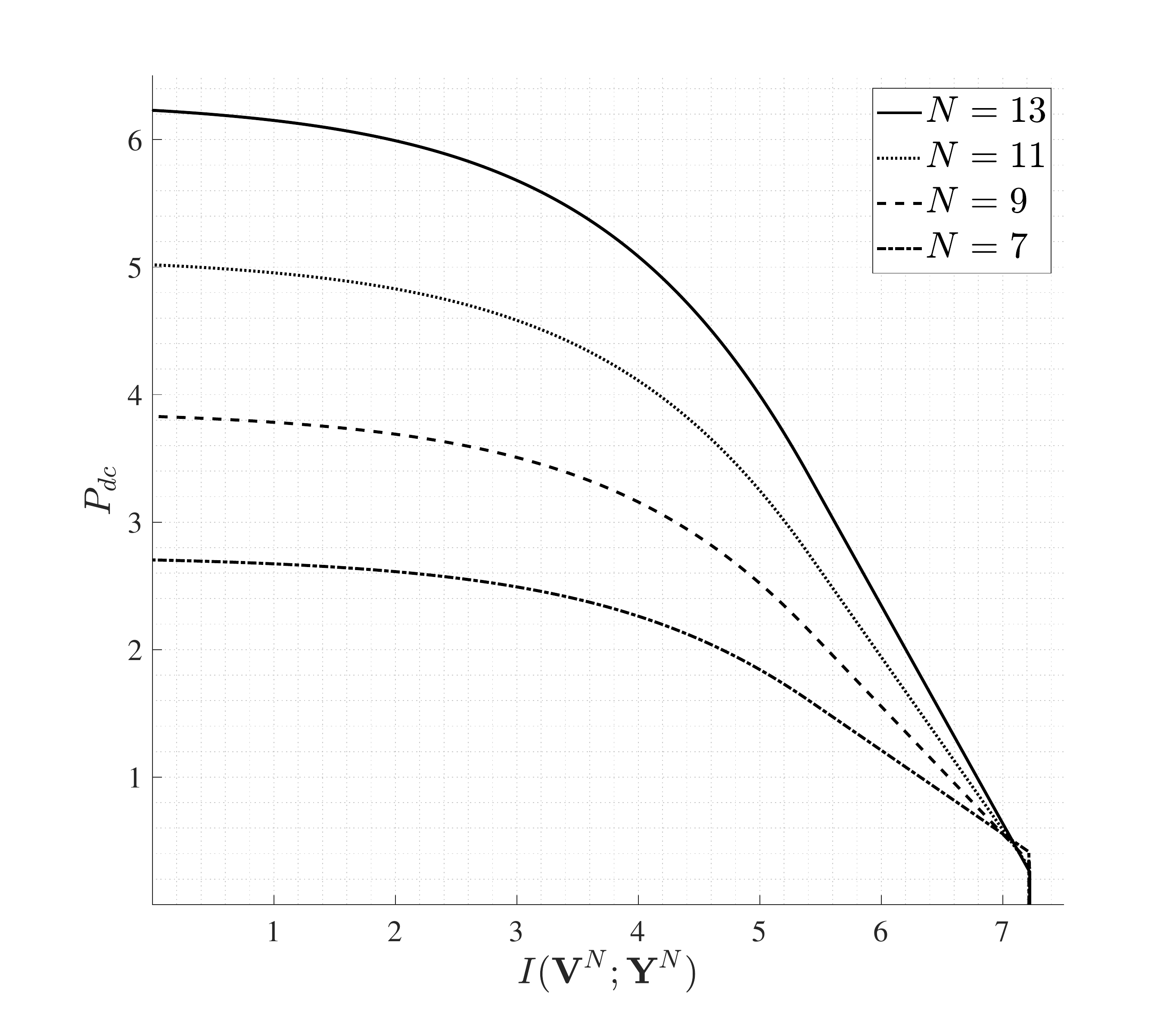}
\caption{The optimized RP regions corresponding to nonezero mean Gaussian inputs for $N=3,~5,~7,~9$ with an average power constraint $P_a=1$ and noise variance $\sigma_w^2=0.1$.}\label{Fig_10}
\par\end{centering}
\vspace{0mm}
\end{figure}

\begin{figure}
\begin{centering}
\includegraphics[scale=0.25]{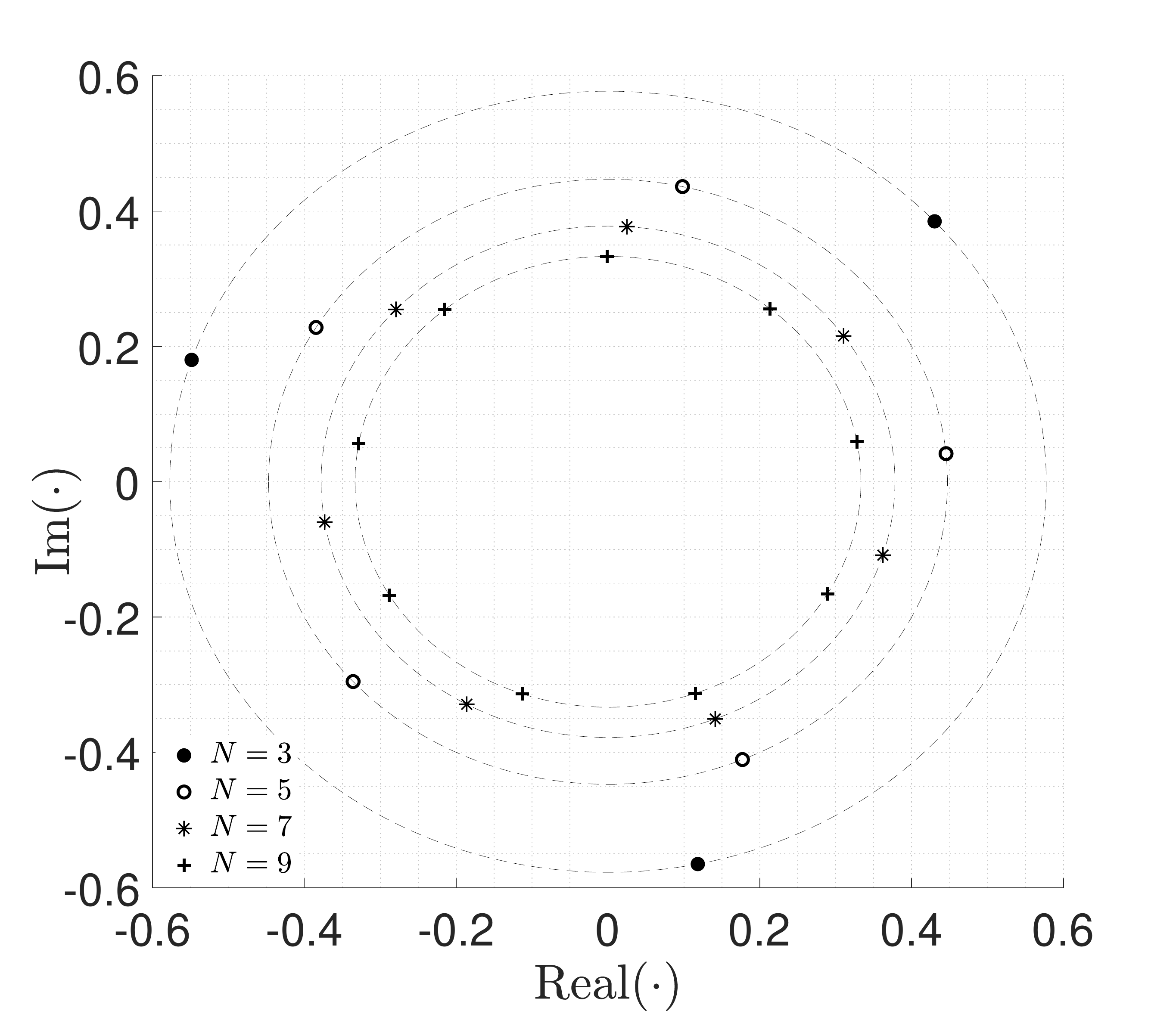}
\caption{The numerically optimized inputs for WPT (under flat fading assumption for the channel) for $N=3,~5,~7,~9$ with an average power constraint $P_a=1$ and noise variance $\sigma_w^2=0.1$.}\label{Fig_11}
\par\end{centering}
\vspace{0mm}
\end{figure}

\begin{rem}
In Figure \ref{Fig_10} (using the optimization algorithm, explained earlier in Algorithm \ref{euclid}) the RP regions are obtained for $N=7,~9,~11,~13$. It is observed that the delivered power at the receiver is increased by the number of the c-subchannels $N$. This is due to the presence of input moments (higher than 2) in the delivered power in (\ref{eqn_39}), and is inline with observations made in \cite{Clerckx_Bayguzina_2016,Clerckx_2016}\footnote{We note that, in practical implementations, this observation (increasing delivered power with $N$) cannot be valid for all $N$, and the delivered power is saturated after some $N$. This is due to the diode breakdown effect, which has not been considered in our model (\ref{eqn_2}) due to small signal analysis. This is further discussed in \cite{Clerckx_2016}.}.

As another interesting observation, in Figure \ref{Fig_11}, the numerically optimized inputs for WPT (under the assumption of flat fading for the channel) are illustrated for $N=3,~5,~7,~9$. As mentioned in Algorithm \ref{euclid}, for each $N$, the optimization is run for many times, each time fed with a randomly generated starting point. In Figure \ref{Fig_11}, the optimized inputs for WPT purposes (zero variance inputs) are illustrated. The phases of the mean on different c-subchannels are also equally spaced.
\end{rem}

\section{Conclusion}\label{Sec:Conc}
In this paper, we studied SWIPT signalling for frequency-selective channels under transmit average power and receiver delivered power constraints. We considered an approximation for the nonlinear EH, which is based on truncation (up to fourth moment) of the Taylor expansion of the rectenna's diode characteristic function. For independent input distributions on different r-subchannels and iid inputs on each r-subchannel, we obtained the delivered power in terms of the system baseband parameters, which demonstrates the dependency of the delivered power on the mean as well as higher moments of the channel input distribution. Assuming that the transmitter is constrained to utilize Gaussian distributions, we show that in general non-zero mean Gaussian inputs attain a larger RP region compared to their zero mean counterparts. As a special scenario, for zero mean Gaussian inputs, we obtained the conditions for optimal power allocation on different r-subchannels. Using numerical optimization, it is observed that optimized non-zero mean inputs (with asymmetric power allocation in each c-subchannel) achieve larger RP region, compared to their optimized  zero mean as well as non-zero mean (with symmetric power allocation in each c-subchannel \cite{Clerckx_2016}) counterparts.

\appendix
\section{Proof of the Proposition \ref{Lemma1}} \label{app:2}
In the following, we obtain the baseband equivalent of (\ref{eqn_2}). Considering first the term $\mathbb{E}\mathcal{E}[Y_{\text{rf}}(t)^2]$, we have
\begin{align}\nonumber
  \mathbb{E}\mathcal{E}[Y_{\text{rf}}(t)^2] &=\frac{1}{2} \mathbb{E}\mathcal{E}\left[\left(Y(t)e^{jf_ct}+Y^{*}(t)e^{-jf_ct}\right)^2\right] \\\label{eqn_25}
  &=\mathbb{E}\mathcal{E}\left[|Y(t)|^2\right]\\\nonumber
  &=\mathbb{E}\mathcal{E}\left[\sum\limits_{n,m}Y[n]Y[m]^* \mathrm{sinc}(f_wt-n)\mathrm{sinc}(f_wt-m)\right]\\\nonumber
  &=\sum\limits_{n,m}\mathbb{E}\left[Y[n]Y[m]^*\right] \mathcal{E}\left[\mathrm{sinc}(f_wt-n)\mathrm{sinc}(f_wt-m)\right]\\\label{eqn_26}
  &=\lim_{T\rightarrow\infty }\frac{1}{Tf_w}\left(\sum\limits_{n}\mathbb{E}\left[|Y[n]|^2\right]\right)\\\label{eqn_9}
  &\simeq \lim_{T\rightarrow\infty }\frac{1}{Tf_w}\left(\sum\limits_{s=1}^{N_b} \sum\limits_{n=(s-1)(L+N-1)+L}^{s(L+N-1)}\mathbb{E}\left[|Y[n]|^2\right]\right)\\\label{eqn_10}
  &=\lim_{T\rightarrow\infty }\frac{1}{Tf_w}\left(\sum\limits_{s=1}^{N_b}\sum\limits_{l=(s-1)N}^{sN-1}\mathbb{E}\left[|Y_l|^2\right]\right)\\\nonumber
  &=\lim_{T\rightarrow\infty }\frac{1}{Tf_w}\sum\limits_{s=1}^{N_b}\sum\limits_{l=(s-1)N}^{sN-1}\left(|h_l|^2 P_l+\sigma_w^2\right)\\\label{eqn_44}
  &=\sum\limits_{l=0}^{N-1}\left(|h_l|^2 P_l+\sigma_w^2\right),
\end{align}
where (\ref{eqn_25}) is due to $\mathcal{E}[Y(t)^{2}e^{2jf_ct}]=\mathcal{E}[Y(t)^{*2}e^{-2jf_ct}]=0$. (\ref{eqn_26}) is due to \[\int_{-\infty}^{\infty}\mathrm{sinc}(f_wt-n)\mathrm{sinc}(f_wt-m)dt=\frac{1}{f_w}\delta(m-n).\]  In (\ref{eqn_9}), $s$ is the OFDM symbol index and $N_b=f_wT/(N+L)$. Note that, in (\ref{eqn_9}) we neglect the delivered power due to utilizing the cyclic prefix (CP). In (\ref{eqn_10}) the result is due to Parseval's theorem.

Next, considering the second term in (\ref{eqn_2}), i.e., $\mathbb{E}\mathcal{E}[Y_{\text{rf}}(t)^4]$, we have
\begin{align}\nonumber
  \mathbb{E}\mathcal{E}[Y_{\text{rf}}(t)^4] &=\frac{1}{4} \mathbb{E}\mathcal{E}\left[4|Y(t)|^4+(Y(t)^2e^{j2f_ct}+{Y(t)^*}^2e^{-j2f_ct})^2\right. \\\nonumber
  &\left.~~+4|Y(t)|^2(Y(t)^2e^{j2f_ct}+{Y(t)^*}^2e^{-j2f_ct})\right] \\\label{eqn_15}
  &=\frac{3}{2}\mathbb{E}\mathcal{E}\left[|Y(t)|^4\right].
\end{align}
Note that the signal $|Y(t)|^2$ is real with bandwidth $(-f_w,f_w)$. Therefore, (\ref{eqn_15}) can be rewritten as
\begin{align}\label{eqn_13}
 \frac{3}{2}\mathbb{E}\mathcal{E}\left[|Y(t)|^4\right]&=\lim_{T\rightarrow\infty} \frac{3}{4Tf_w}\sum\limits_{n} \left(\mathbb{E}\left[|Y[n]|^4\right]+\mathbb{E}\left[|\tilde{Y}[n]|^4\right]\right)\\\nonumber
 &=\lim_{T\rightarrow\infty} \frac{3}{4Tf_w}\sum\limits_{s=1}^{N_b}\sum\limits_{n=(s-1)(L+N-1)+L}^{s(L+N-1)} \left(\mathbb{E}\left[|Y[n]|^4\right]+\mathbb{E}\left[|\tilde{Y}[n]|^4\right]\right)\\\nonumber
 &\simeq \frac{3}{4} \sum_{n=L}^{N+L-1} \left(\mathbb{E}\left[|Y[n]|^4\right]+\mathbb{E}\left[|\tilde{Y}[n]|^4\right]\right),
\end{align}
where $Y[n]$ and $\tilde{Y}[n]$ in (\ref{eqn_13}) are the samples of $Y(t)$ taken at times $t=\frac{2n}{2f_w}$ and $t=\frac{2n+1}{2f_w}$, respectively, i.e., $Y[n]\triangleq Y(\frac{2n}{2f_w})$ and $\tilde{Y}[n]\triangleq Y(\frac{2n+1}{2f_w})$. In the following, we analyze $Y[n]$ and $\tilde{Y}[n]$, separately.

\subsection{Samples of $Y(t)$ at times $t=\frac{2n}{2f_w}$:}
First considering $Y[n]$, (in one OFDM symbol) we have
\begin{align}\nonumber
  \sum_{n=L}^{L+N-1}\mathbb{E}\left[|Y[n]|^4\right]&=\sum_{n=L}^{L+N-1}\mathbb{E}\left[(|Y[n]|^2)^2\right]\\\label{eqn_17}
  &=\frac{1}{N}\sum_{k=0}^{N-1}\mathbb{E}\left[|Y_{k} \otimes Y_{(-k)_N}^*|^2\right]\\\label{eqn_18}
  &=\frac{1}{N}\sum_{k=0}^{N-1}\sum_{l=0}^{N-1}\sum_{m=0}^{N-1} \mathbb{E}\left[Y_{l}Y_{(l-k)_N}^*Y_{m}^*Y_{(m-k)_N}\right],
\end{align}
where (\ref{eqn_17}) is due to Parseval's theorem and convolution property of DFT. Define $F_l\triangleq h_lV_l$. We have $Y_{l}=F_l+W_l$. By expanding (\ref{eqn_18}) we have\footnote{In (\ref{eqn_19}) and (\ref{eqn_18}), in each term, the first, second, third and fourth letter have the subscript indices $l,(l-k)_N,m,(m-k)_N$, respectively. Also the second and the third letter bear a conjugate sign. We have removed the indices as well as the conjugates for clarity.}
\begin{align}\nonumber
\mathbb{E}\big[Y_{l}Y_{(l-k)_N}^{*}&Y_{m}^*Y_{(m-k)_N}\big]= \mathbb{E}\left[FFFF+FFFW+FFWF+FFWW\right.\\\nonumber
  &+FWFF+FWFW+FWWF+FWWW\\\nonumber
  &+WFFF+WFFW+WFWF+WFWW\\\label{eqn_19}
  &\left.+WWFF+WWFW+WWWF+WWWW\right]\\\label{eqn_20}
  &= \mathbb{E}\left[FFFF+FFWW+FWFW+WFWF+WWFF+WWWW\right].
\end{align}
In the following, we calculate each of the terms in (\ref{eqn_20}) (Note that since the noise is CSCG, we have $\mathbb{E}[|W|^4]=2\sigma_w^4$ and $\mathbb{E}[{W^*}^2]=\mathbb{E}[W^2]=0$)
\begin{align}\nonumber
\sum_{k,l,m=0}^{N-1}\mathbb{E}\left[W_lW_{(l-k)_N}^*W_{m}^*W_{(m-k)_N}\right]&=\sum_{k,l,m=0}^{N-1} \delta_{(l-m)_N}(\delta_k2\sigma_w^4+(1-\delta_k)\sigma_w^4) +(1-\delta_{(l-m)_N})\delta_{k}\sigma_w^4\\\nonumber
&=\sum_{l=0}^{N-1}2\sigma_w^4+\sum_{l=0}^{N-1}\sum_{k=1}^{N-1}\sigma_w^4+\sum_{l=0}^{N-1}\sum_{m=0,m\neq l}^{N-1}\sigma_w^4\\\label{eqn_14}
&=\sum_{l=0}^{N-1}2N^2\sigma_w^4,\\\nonumber
\sum_{k,l,m=0}^{N-1}\mathbb{E}\left[W_lW_{(l-k)_N}^*F_{m}^*F_{(m-k)_N}\right]&=\sum_{k,l,m=0}^{N-1}\delta_k\sigma_w^2 |h_m|^2\mathbb{E}[|V_m|^2]\\\label{eqn_16}
&=\sum_{l=0}^{N-1}N\sigma_w^2 |h_l|^2P_l,\\\nonumber
\sum_{k,l,m=0}^{N-1}\mathbb{E}\left[W_lF_{(l-k)_N}^*W_{m}^*F_{(m-k)_N}\right] &=\sum_{k,l,m=0}^{N-1}\delta_{(l-m)_N}\sigma_w^2|h_{(m-k)_N}|^2\mathbb{E}\left[|V_{(m-k)_N}|^2\right]\\\nonumber
&=\sum_{k,m}\sigma_w^2|h_{(m-k)_N}|^2\mathbb{E}\left[|V_{(m-k)_N}|^2\right]\\\label{eqn_11}
&=\sum_{l=0}^{N-1}N\sigma_w^2|h_l|^2P_l,
\end{align}
where (\ref{eqn_11}) is due to the property of circular convolution. For the other terms we have
\begin{align}\nonumber
\sum_{k,l,m=0}^{N-1}\mathbb{E}\left[F_lW_{(l-k)_N}^*F_{m}^*W_{(m-k)_N}\right]&= \sum_{k,l,m=0}^{N-1}\delta_{(l-m)_N}\sigma_w^2|h_m|^2P_m\\\label{eqn_27}
&=\sum_{l=0}^{N-1}N\sigma_w^2|h_l|^2P_l,\\\nonumber
\sum_{k,l,m=0}^{N-1}\mathbb{E}\left[F_lF_{(l-k)_N}^*W_{m}^*W_{(m-k)_N}\right]&=\sum_{k,l,m=0}^{N-1}\delta_k\sigma_w^2 |h_l|^2\mathbb{E}[|V_l|^2]\\\label{eqn_28}
&=\sum_{l=0}^{N-1}N\sigma_w^2 |h_{l}|^2P_l,
\end{align}
\begin{align}
&\sum_{k,l,m=0}^{N-1}\mathbb{E}\left[F_lF_{(l-k)_N}^*F_{m}^*F_{(m-k)_N}\right]=\sum_{l,m,k=0}^{N-1}\Big\{(1-\delta_{(l-m)_N})(1-\delta_k)E_{l,m,k}\\\nonumber
&+\delta_{(l-m)_N}(1-\delta_k)|h_l|^2|h_{(l-k)_N}|^2P_lP_{(l-k)_N}+\delta_{(l-m)_N}\delta_k|h_l|^4Q_l+(1-\delta_{(l-m)_N})\delta_k|h_l|^2|h_m|^2P_lP_m\Big\}\\\nonumber
&=\sum_{l,m,k=0}^{N-1}(1-\delta_{(l-m)_N})(1-\delta_k)E_{l,m,k}+\sum_{k=1}^{N-1}\sum_{l=0}^{N-1}|h_l|^2|h_{(l-k)_N}|^2P_lP_{(l-k)_N}\\\nonumber
&+\sum_{l=0}^{N-1}|h_l|^4Q_l+\sum_{l=0}^{N-1}\sum_{m=0,m\neq l}^{N-1}|h_l|^2|h_m|^2P_lP_m\\\nonumber
&=\sum_{l,m,k=0}^{N-1}(1-\delta_{(l-m)_N})(1-\delta_k)E_{l,m,k}+2\sum_{l=0}^{N-2}\sum_{m=l+1}^{N-1}|h_l|^2|h_m|^2P_lP_m\\\nonumber
&+\sum_{l=0}^{N-1}|h_l|^4Q_l+2\sum_{l=0}^{N-2}\sum_{m=l+1}^{N-1}|h_l|^2|h_m|^2P_lP_m\\\label{eqn_29}
&=\sum_{l,m,k=0}^{N-1}(1-\delta_{(l-m)_N})(1-\delta_k)E_{l,m,k}+4\sum_{l=0}^{N-1}\delta^l_{N-1}\sum_{m=l+1}^{N-1}|h_l|^2|h_m|^2P_lP_m+\sum_{l=0}^{N-1}|h_l|^4Q_l.
\end{align}
For $E_{l,m,k}$ we have $l\neq m,~k\neq 0$. Hence, we also have $l\neq (l-k)_N,~m\neq (m-k)_N$. Four different situations occur. We have
\begin{align}\nonumber
E_{l,m,k}=&\left\{\begin{array}{ll}
\mathbb{E}[F_l^2]\mathbb{E}[{F_m^*}^2] &~l=(m-k)_N,m=(l-k)_N\\
\mathbb{E}[F_l^2]\mathbb{E}[F_m^*]\mathbb{E}[F_{(l-k)_N}^*]&~l=(m-k)_N,m\neq(l-k)_N\\
\mathbb{E}[{F_m^*}^2]\mathbb{E}[F_l]\mathbb{E}[F_{(m-k)_N}] &~l\neq (m-k)_N ,m=(l-k)_N\\
\mathbb{E}[F_l]\mathbb{E}[F_m^*]\mathbb{E}[F_{(l-k)_N}^*]\mathbb{E}[F_{(m-k)_N}]&~l\neq (m-k)_N ,m\neq(l-k)_N
\end{array}\right.\\\label{eqn_30}
=&\left\{\begin{array}{ll}
h_l^2{h_m^*}^2\bar{P_l}\bar{P_m^*}&~l=(m-k)_N,m=(l-k)_N \\
h_l^2h_m^*h_{(l-k)_N}^*\bar{P_l}\mu_m^*\mu_{(l-k)_N}^*&~l=(m-k)_N,m\neq(l-k)_N\\
h_m^{{*}2}h_lh_{(m-k)_N}\bar{P_m^*}\mu_l\mu_{(m-k)_N}&~l\neq (m-k)_N ,m=(l-k)_N\\
h_lh_m^*h_{(l-k)_N}^*h_{(m-k)_N}\mu_{l}\mu_{m}^*\mu_{(l-k)_N}^*\mu_{(m-k)_N}&~l\neq (m-k)_N ,m\neq(l-k)_N
\end{array}
\right..
\end{align}

Substituting the terms in (\ref{eqn_30}) in $E_{l,m,k}$ in (\ref{eqn_29}), we have
\begin{align}\nonumber
&\sum_{k,l,m=0}^{N-1}\mathbb{E}\left[F_lF_{(l-k)_N}^*F_{m}^*F_{(m-k)_N}\right]=\sum_{\substack{k\neq 0,l\neq m\\l=(m-k)_N\\m=(l-k)_N}}\!\!\!\!\!h_l^2{h_m^*}^2\bar{P_l}\bar{P_m^*}\\\nonumber
&\!\!\!+\sum_{\substack{k\neq 0,l\neq m\\l=(m-k)_N\\m\neq (l-k)_N}}\!\!\!\!\!h_l^2h_m^*h_{(l-k)_N}^*\bar{P_l}\mu_m^*\mu_{(l-k)_N}^*+\sum_{\substack{k\neq 0,l\neq m\\l\neq (m-k)_N\\m=(l-k)_N}}\!\!\!\!\!h_m^{{*}2}h_lh_{(m-k)_N}\bar{P_m^*}\mu_l\mu_{(m-k)_N}\\\label{eqn_36}
&\!\!\!+\sum_{\substack{k\neq 0,l\neq m\\l\neq (m-k)_N\\m\neq (l-k)_N}}\!\!\!\!\!h_lh_m^*h_{(l-k)_N}^*h_{(m-k)_N}\mu_{l}\mu_{m}^*\mu_{(l-k)_N}^*\mu_{(m-k)_N}+4\sum_{l=0}^{N-1}\delta^l_{N-1}\sum_{m=l+1}^{N-1}|h_l|^2|h_m|^2P_lP_m+\sum_{l=0}^{N-1}|h_l|^4Q_l.
\end{align}

For the first term in the RHS of (\ref{eqn_36}), it is verified that (we recall that $N$ is odd)
\begin{align}\label{M1}
  \sum_{\substack{k\neq 0,l\neq m\\l=(m-k)_N\\m=(l-k)_N}}\!\!\!\!\!h_l^2{h_m^*}^2\bar{P_l}\bar{P_m^*}&=0.
\end{align}
This is because from $l=(m-k)_N$ and $m=(l-k)_N$, we have $l=(l-2k)_N$, which results in $k=0,N/2$. Noting that $N$ is odd and $k\neq 0$, there is no such integer. For the second and third terms in the RHS of (\ref{eqn_36}), we have $l=(m-k)_N$ and $m=(l-k)_N$, respectively. Therefore, we obtain
\begin{align}\label{eqn_31}
  \sum_{\substack{k\neq 0,l\neq m\\l=(m-k)_N\\m\neq(l-k)_N}}\!\!\!\!\!h_l^2\bar{P_l}h_m^*h_{(l-k)_N}^*\mu_m^*\mu_{(l-k)_N}^*&= \sum_{l=0}^{N-1}\sum_{k=1}^{N-1}h_l^2\bar{P}_lh_{(l+k)_N}^*h_{(l-k)_N}^*\mu_{(l+k)_N}^*\mu_{(l-k)_N}^*,\\\label{eqn_32}
  \sum_{\substack{k\neq 0,l\neq m\\l\neq(m-k)_N\\m=(l-k)_N}}\!\!\!\!\!h_m^{{*}2}\bar{P}_m^*h_lh_{(m-k)_N}\mu_l\mu_{(m-k)_N}&= \sum_{l=0}^{N-1}\sum_{k=1}^{N-1}h_l^{*2}\bar{P}_l^*h_{(l+k)_N}h_{(l-k)_N}\mu_{(l+k)_N}\mu_{(l-k)_N}.
\end{align}
Since the terms in (\ref{eqn_31}) and (\ref{eqn_32}) are conjugate of each other, therefore by adding the two, we get
\begin{align}\label{M2}
  \text{Term in (\ref{eqn_31})}+\text{Term in (\ref{eqn_32})}=2\sum_{l=0}^{N-1}\sum_{k=1}^{N-1}\Re\{h_l^2\bar{P}_lh_{(l+k)_N}^*h_{(l-k)_N}^*\mu_{(l+k)_N}^*\mu_{(l-k)_N}^*\}.
\end{align}

For the fourth term in the RHS of (\ref{eqn_36}), note that since $k\neq 0$, $m\neq l$, $l\neq (m-k)_N$ and $m\neq (l-k)_N$, overall we have $N(N-1)(N-3)$ terms. Before simplifying this term, we mention the following two points\footnote{We recall that $N$ is odd. Similar discussion can be used to simplify the results for even $N$. However, since calculations for odd $N$ is easier to follow, we opt to bring the discussion for odd $N$ only.}:

\begin{figure}[t!]
\begin{centering}
\begin{subfigure}
  \par\hfill
  \includegraphics[scale=0.29]{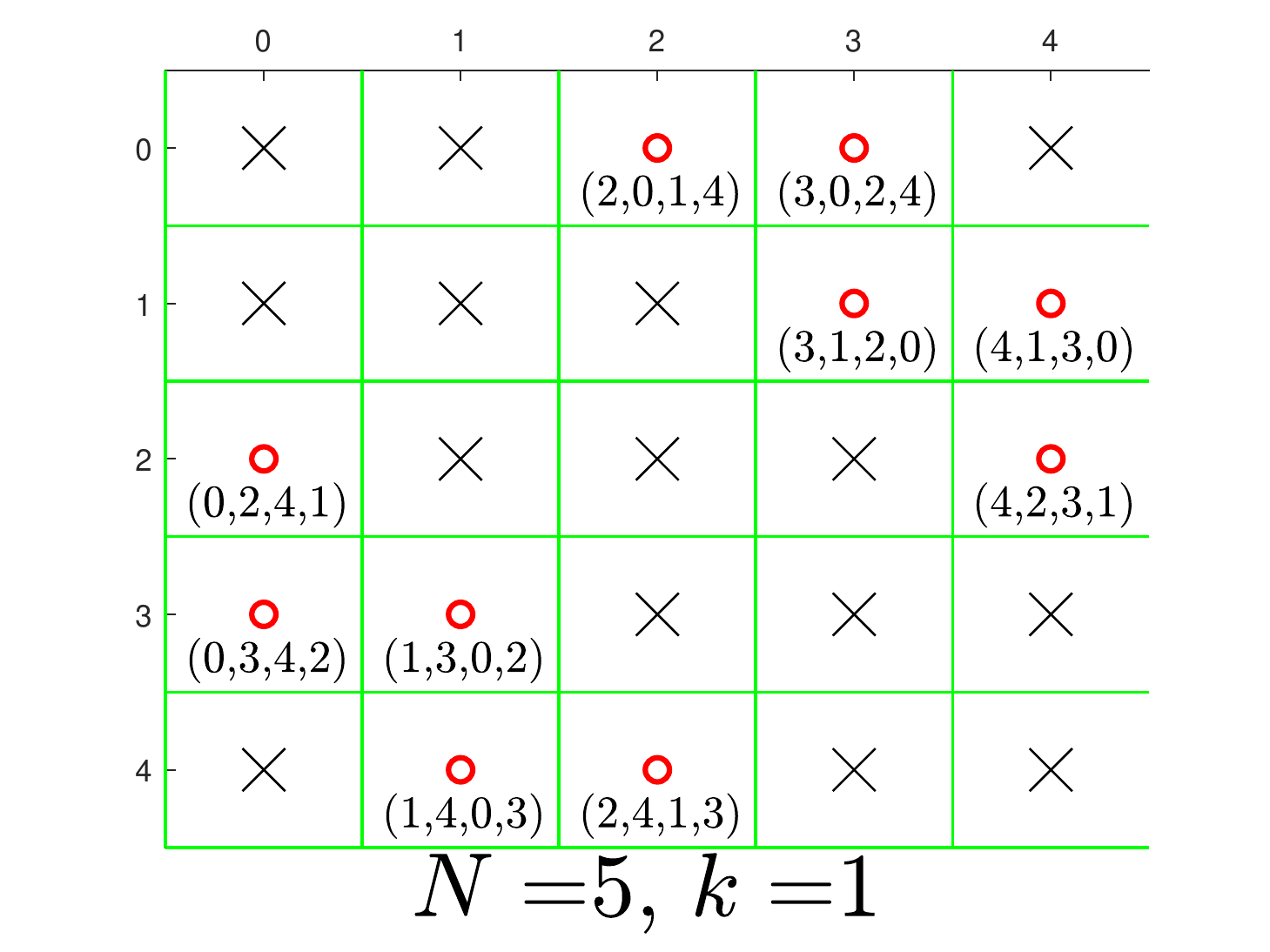}
\end{subfigure}
\begin{subfigure}
  \par\hfill
  \includegraphics[scale=0.29]{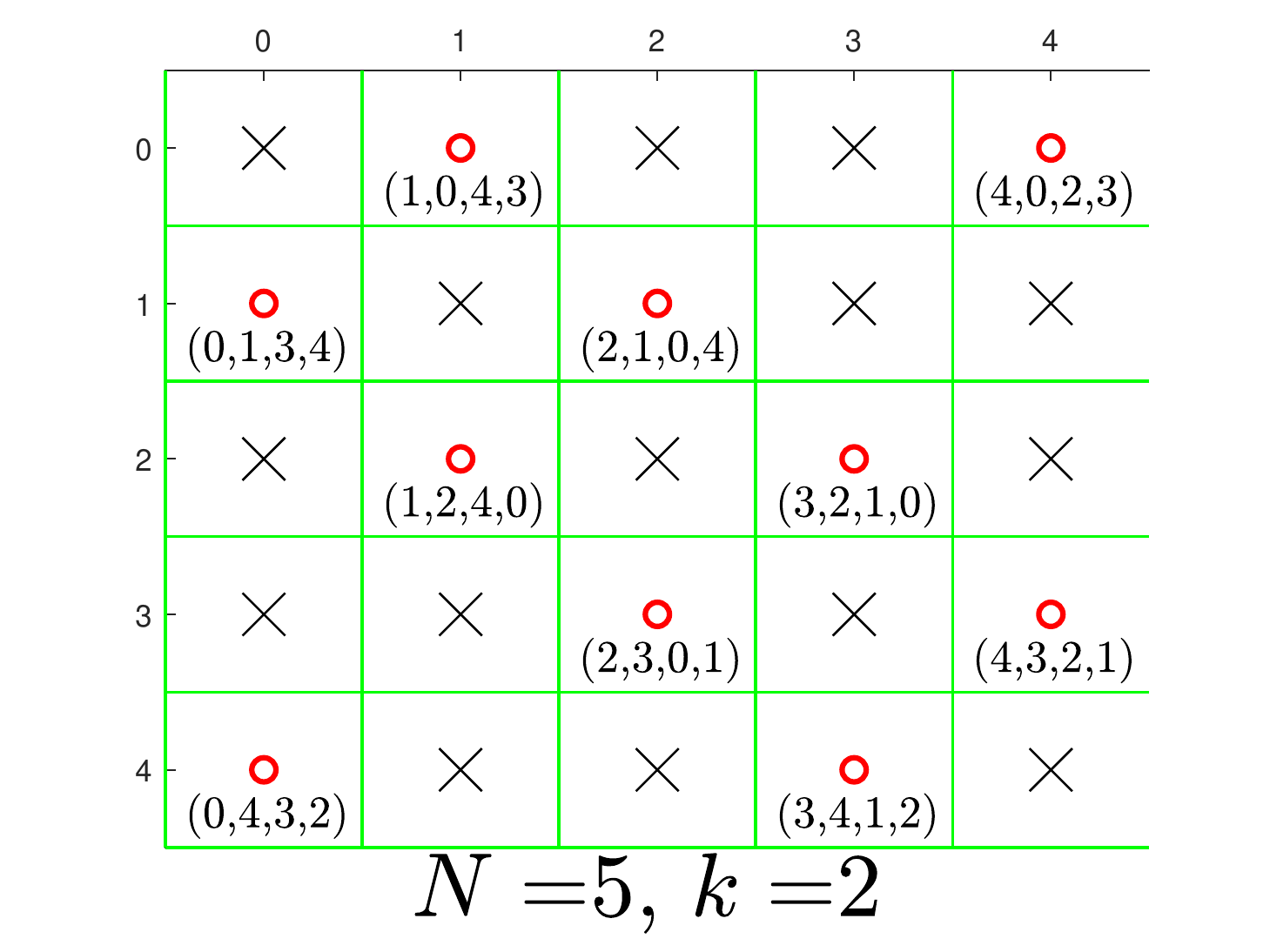}
\end{subfigure}
\begin{subfigure}
  \par\hfill
  \includegraphics[scale=0.29]{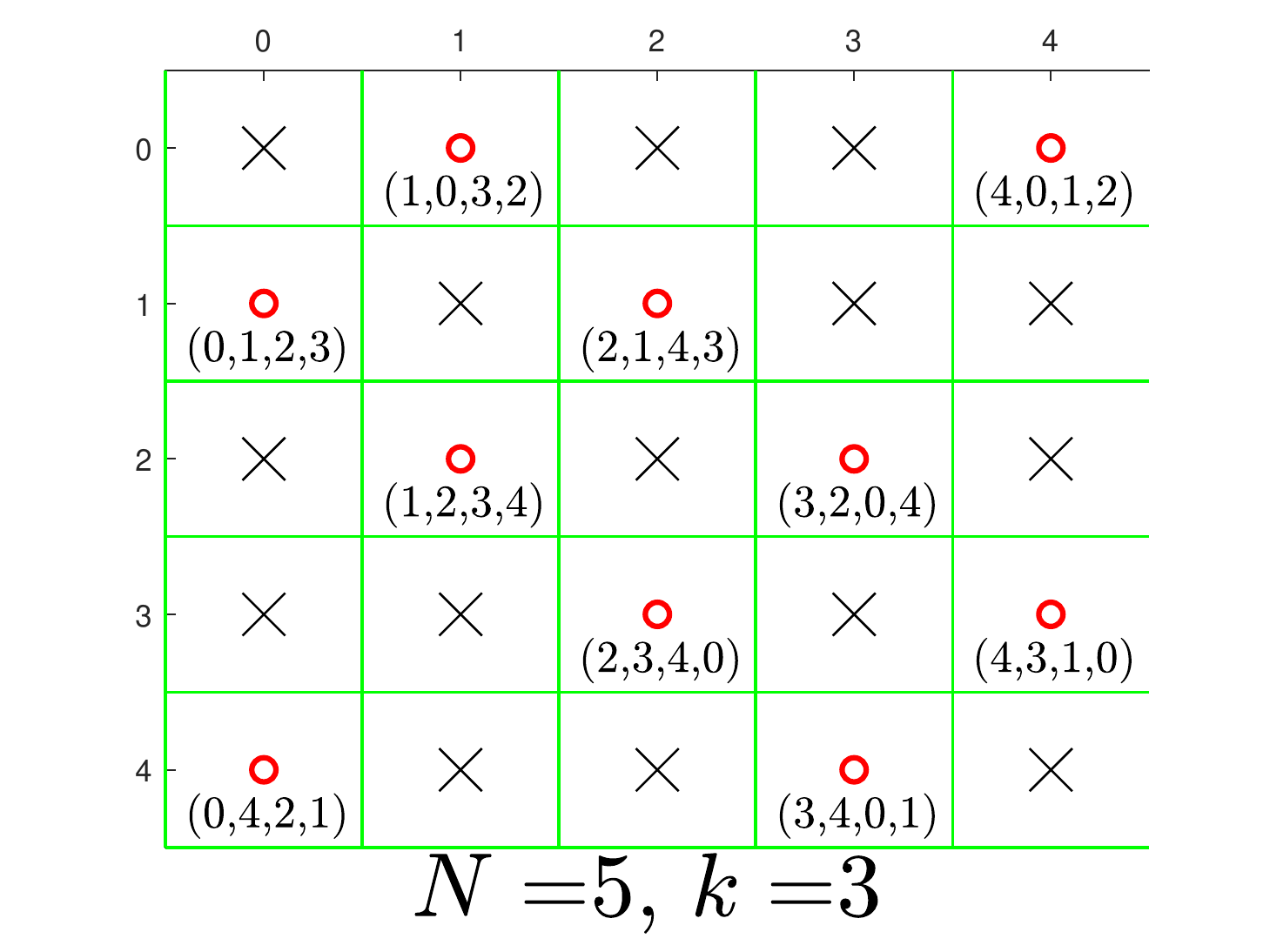}
\end{subfigure}
\begin{subfigure}
  \par\hfill
  \includegraphics[scale=0.29]{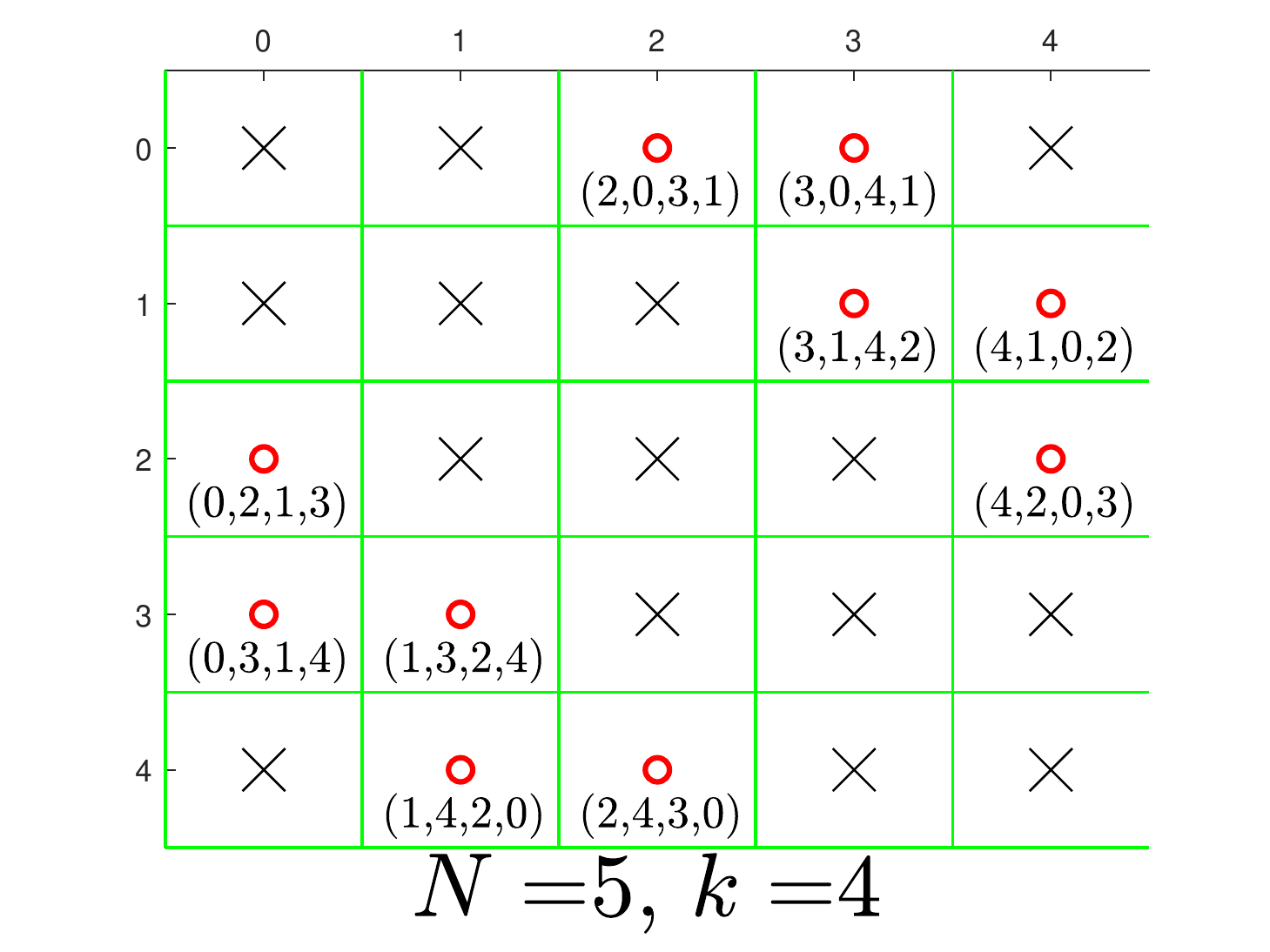}
\end{subfigure}
\caption{Illustration of the admissible indices for the fourth term in the RHS of (\ref{eqn_36}) for $N=5$. The red dots are the admissible triplet indices $(l,m,k)$, whereas the others are not. The quadruplet indices under each red dot represents the numbers $(l,m,(l-k)_N,(m-k)_N)$ corresponding to the indices of each term for the fourth summation in the RHS of (\ref{eqn_36}). Note the diagonal symmetry for a given $k$, as well as the symmetry between ($k=1$ and $k=4$) or ($k=2$ and $k=3$).}\label{Fig_5}
\end{centering}
\end{figure}

\begin{itemize}
 \item \textbf{Symmetry with respect to the index $k$}: Note that for the fourth term in the RHS of (\ref{eqn_36}) for $k=i,~i=1,\ldots,\frac{N-1}{2}$, if the term $a_la_m^{*}a_{(l-i)_N}^{*}a_{(m-i)_N}$\footnote{For brevity we define $a_l\triangleq h_l\mu_l$.} is admissible (the red circles in Figure \ref{Fig_5}), then it is easy to verify that there exists the same term in the ordering of $a_{(m-i)_N}a_{(l-i)_N}^{*}a_m^{*}a_l$\footnote{Note the term $a_la_m^{*}a_{(l-k)_N}^{*}a_{(m-k)_N}$ is the same if we replace the first with fourth and the second with the third element.} with the corresponding index $k=N-i$. As an example for clarification, this can also be verified by considering the tables corresponding to $k=1$ and $k=4$ (or $k=2$ and $k=3$) together in Figure \ref{Fig_5}. Therefore, due to this property it is enough to consider the summation over the indices $k=1:\frac{N-1}{2}$ multiplied by a factor two.
 \item \textbf{Diagonal symmetry between $m$ and $l$ for a given $k$}: Note that for a fixed index $k$, the admissible term $a_la_m^*a_{(l-k)_N}^*a_{(m-k)_N}$ is conjugate of the admissible term $a_ma_l^*a_{(m-k)_N}^*a_{(l-k)_N}$. This accordingly shows in the table of indices corresponding to the given index $k$ (similar to each one of the tables in Figure \ref{Fig_5}), that the upper-diagonal admissible set of elements are conjugate of the lower-diagonal admissible ones. Since we consider the summation of these elements, it is enough to run the summation over the twice of real of lower-diagonal set of admissible elements.
\end{itemize}

Therefore, according to the aforementioned points, the fourth term on the RHS of (\ref{eqn_36}) is simplified as
\begin{align}\label{eqn_35}
  \sum_{\substack{k\neq 0,l\neq m\\
  l\neq(m-k)_N\\m\neq(l-k)_N}}\!\!\!\!\!a_la_m^*a_{(l-k)_N}^*a_{(m-k)_N}&= 4 \sum_{k=1}^{\frac{N-1}{2}}\sum_{l=0}^{N-2}\sum_{\substack{m=l+1\\m\neq(l+k)_N\\m\neq (l-k)_N}}^{N-1}\Re\{a_la_m^*a_{(l-k)_N}^*a_{(m-k)_N}\}
\end{align}


Substituting (\ref{M1}), (\ref{M2}) and (\ref{eqn_35}) in (\ref{eqn_36}), we have
\begin{align}\nonumber
\sum_{k,l,m}\mathbb{E}\left[F_lF_{(l-k)_N}^*F_{m}^*F_{(m-k)_N}\right]&= 4\sum_{l=0}^{N-1}\delta^l_{N-1}\sum_{m=l+1}^{N-1}|h_l|^2|h_m|^2P_lP_m+\sum_{l=0}^{N-1}|h_l|^4Q_l\\\nonumber
&+2\sum_{l=0}^{N-1}\sum_{k=1}^{N-1}\Re\{h_l^2\bar{P}_lh_{(l+k)_N}^*h_{(l-k)_N}^*\mu_{(l+k)_N}^*\mu_{(l-k)_N}^*\}\\\label{eqn_38}
&+4 \sum_{l=0}^{N-2}\sum_{k=1}^{\frac{N-1}{2}}\sum_{\substack{m=l+1\\m\neq(l+k)_N\\m\neq (l-k)_N}}^{N-1}\Re\{h_lh_m^*h_{(l-k)_N}^*h_{(m-k)_N}\mu_{l}\mu_{m}^*\mu_{(l-k)_N}^*\mu_{(m-k)_N}\}
\end{align}

\subsection{Samples of $Y(t)$ at times $t=\frac{2n+1}{2f_w}$:}\label{Sec:Channel_u}
The continuous baseband received signal $Y(t)$ can be written as \cite[Chapter 2]{Tse:book}
\begin{align}
  Y(t)&=\sum_{k}X[k]\sum_{i}a_i^b(t)\mathrm{sinc}(f_wt-f_w\tau_i(t)-k)+W(t).
\end{align}

Considering the samples at times $t=\frac{2n+1}{f_w}$ for integer $n$, we have
\begin{align}
  Y\left(\frac{2n+1}{2f_w}\right)&=\sum_{k}X[k]\sum_{i}a_i^b\left(\frac{2n+1}{2f_w}\right)\mathrm{sinc} \left(\frac{2n+1}{2}-f_w\tau_i\left(\frac{2n+1}{2f_w}\right)-k\right)+W\left(\frac{2f_w+1}{2f_w}\right)\\
  &=\sum_{k}X[k]\sum_{i}a_i^b\left(\frac{2n+1}{2f_c}\right)\mathrm{sinc} \left(n+\frac{1}{2}-f_w\tau_i\left(\frac{2n+1}{2f_w}\right)-k\right)+W\left(\frac{n}{f_w}+\frac{1}{2f_w}\right)\\
  &=\sum_{l}X[n-l]\sum_{i}a_i^b\left(\frac{2n+1}{2f_w}\right)\mathrm{sinc} \left(l+\frac{1}{2}-f_w\tau_i\left(\frac{2n+1}{2f_w}\right)\right)+W\left(\frac{n}{f_w}+\frac{1}{2f_w}\right)\\
  &=\sum_{l=0}^{L-1}X[n-l]\tilde{h}^{u}[l]+W\left(\frac{n}{f_w}+\frac{1}{2f_w}\right),
\end{align}
where $\tilde{h}_{l}^u\triangleq\sum_{i}a_i^b\left(\frac{2n+1}{2f_w}\right)\mathrm{sinc} \left(l+\frac{1}{2}-f_w\tau_i\left(\frac{2n+1}{2f_w}\right)\right)$ stands for the samples (in the time domain) of the channel at times $t=\frac{2n+1}{2f_w}$. Recalling that $\tilde{Y}[n]\triangleq Y\left(\frac{2n+1}{2f_w}\right)$, we have
\begin{align}
  \sum_{n=L}^{L+N-1}\mathbb{E}\big[\big|\tilde{Y}[n]\big|^4\big]&=\frac{1}{N}\sum_{k=0}^{N-1}\mathbb{E}\left[|\tilde{Y}_{k} \otimes \tilde{Y}_{-k}^*|^2\right]\\
  &=\frac{1}{N}\sum_{k=0}^{N-1}\sum_{l=0}^{N-1}\sum_{m=0}^{N-1} \mathbb{E}\left[\tilde{Y}_{l}\tilde{Y}_{(l-k)_N}^*\tilde{Y}_{m}^*\tilde{Y}_{(m-k)_N}\right].
\end{align}
Similarly to (\ref{eqn_19}), we have\footnote{As in (\ref{eqn_19}) and (\ref{eqn_18}), here in (\ref{eqn_37}) in each term, the second and the third symbols are conjugate with the indices $l,(l-k)_N,m,(m-k)_N$ for the first, second, third and the forth term, respectively. For brevity of representations, we have removed the indices as well as the conjugate.}
\begin{align}\label{eqn_37}
\mathbb{E}\left[\tilde{Y}_{l}\tilde{Y}_{(l-k)_N}^*\tilde{Y}_{m}^*\tilde{Y}_{(m-k)_N}\right]
= \mathbb{E}\left[FFFF+FFWW+FWFW+WFWF+WWFF+WWWW\right].
\end{align}
Following the same steps in (\ref{eqn_14}), (\ref{eqn_16}), (\ref{eqn_11}), (\ref{eqn_27}), (\ref{eqn_28}) and ,(\ref{eqn_38}), for each term in (\ref{eqn_37}) we have
\begin{align}
\sum_{k,l,m}\mathbb{E}\left[W_lW_{(l-k)_N}^*W_{m}^*W_{(m-k)_N}\right]&=\sum_{l=0}^{N-1}2N^2\sigma_w^4,\\
\sum_{k,l,m}\mathbb{E}\left[W_lW_{(l-k)_N}^*F_{m}^*F_{(m-k)_N}\right]&=\sum_{l=0}^{N-1}N\sigma_w^2|h^u_l|^2P_l,\\
\sum_{k,l,m}\mathbb{E}\left[W_lF_{(l-k)_N}^*W_{m}^*F_{(m-k)_N}\right]&=\sum_{l=0}^{N-1}N\sigma_w^2 |h^u_l|^2P_l,\\
\sum_{k,l,m}\mathbb{E}\left[F_lW_{(l-k)_N}^*F_{m}^*W_{(m-k)_N}\right]&=\sum_{l=0}^{N-1}N\sigma_w^2 |h^u_l|^2P_l,\\
\sum_{k,l,m}\mathbb{E}\left[F_lF_{(l-k)_N}^*W_{m}^*W_{(m-k)_N}\right]&=\sum_{l=0}^{N-1}N\sigma_w^2|h^u_l|^2P_l,\\\nonumber
\sum_{k,l,m}\mathbb{E}\left[F_lF_{(l-k)_N}^*F_{m}^*F_{(m-k)_N}\right]&= 4\sum_{l=0}^{N-1}\delta^l_{N-1}\sum_{m=l+1}^{N-1}|h^u_l|^2|h^u_m|^2P_lP_m+\sum_{l=0}^{N-1}|h^u_l|^4Q_l\\\nonumber
&+2\sum_{l=0}^{N-1}\sum_{k=1}^{N-1}\Re\{h_l^{u2}\bar{P}_lh_{(l+k)_N}^{u*}h_{(l-k)_N}^{u*}\mu_{(l+k)_N}^*\mu_{(l-k)_N}^*\}\\
&+4 \sum_{l=0}^{N-2}\sum_{k=1}^{\frac{N-1}{2}}\sum_{\substack{m=l+1\\m\neq(l+k)_N\\m\neq (l-k)_N}}^{N-1}\Re\{h^u_lh_m^{u*}h_{(l-k)_N}^{u*}h^u_{(m-k)_N}\mu_{l}\mu_{m}^*\mu_{(l-k)_N}^*\mu_{(m-k)_N}\}
\end{align}
The result of the lemma is obtained by substituting the terms (\ref{eqn_44}) and the baseband equivalent of (\ref{eqn_15}) in (\ref{eqn_2}) altogether.

\section{KKT conditions for the Lagrangian dual function of problem (\ref{eqn_23})}\label{app_KKT}

By writing the Lagrangian for the problem in (\ref{eqn_23}) we have
\begin{align}\nonumber
  L(\lambda_1,\lambda_2,\zeta_{r0},\zeta_{i0},\cdots,\zeta_{r(N-1)},\zeta_{i(N-1)})&=\sum\limits_{l=0}^{N-1}-c_0\big\{\log (1+a_l\sigma_{lr}^2)+\log(1+a_l\sigma_{li}^2)\big\} \\\nonumber
  &+\lambda_1\left(\sum_{l=0}^{N-1} P_l-P^{a}\right)+\lambda_2\left(P_d-\sum_{l=0}^{N-1} f_{ib}(P_l,\bar{P_l},\mu_l,h_l,N)\right)\\
  &-\sum_{l=0}^{N-1}\zeta_{lr}\sigma_{lr}^2+\zeta_{li}\sigma_{li}^2,
\end{align}
where $\lambda_1, \lambda_2\geq 0$ and $\zeta_{lr},\zeta_{li}\geq 0$ for $l=0,\cdots,N-1$ are Lagrange multipliers. Accordingly, the KKT conditions read as
\begin{subequations}\label{eqn_24}
\begin{align}
&\sum_{l=0}^{N-1}P_l\leq P_a,\\
&\sum_{l=0}^{N-1} f_{ib}(P_l,\bar{P_l},\mu_l,h_l,N)\geq P_d,\\
& \sigma_{lr}^2,\sigma_{li}^2\geq 0,~l=0,...,N-1,\\\label{eqn_22}
&\lambda_1\left(\sum_{l=0}^{N-1}P_l- P_a\right)=0,\\
&\lambda_2\left(P_d-\sum_{l=0}^{N-1} f_{ib}(P_l,\bar{P_l},\mu_l,h_l,N) \right)=0,\\
&\zeta_{lr}\sigma_{lr}^2=\zeta_{li}\sigma_{li}^2= 0,~l=0,...,N-1,\\
&\left\{\begin{array}{ll}\label{eqn_12}
  \frac{-c_1a_l}{1+a_l\sigma_{lr}^2}+\lambda_1-\lambda_2 f_{ib}^{P_{lr}}-\zeta_{lr}=0,~l=0,...,N-1\\
  \frac{-c_1a_l}{1+a_l\sigma_{li}^2}+\lambda_1-\lambda_2 f_{ib}^{P_{li}}-\zeta_{li}=0,~l=0,...,N-1\\
\frac{2c_1a_l \mu_{lr}}{1+a_l\sigma_{lr}^2}- \lambda_2 f_{ib}^{\mu_{lr}}+2\zeta_{lr}\mu_{lr}=0,~l=0,...,N-1\\
\frac{2c_1a_l \mu_{li}}{1+a_l\sigma_{li}^2}- \lambda_2 f_{ib}^{\mu_{li}}+2\zeta_{li}\mu_{li}=0,~l=0,...,N-1
\end{array}\right.,
\end{align}
\end{subequations}
where $c_1=c_0\log e$, and
\begin{align}
\left\{\begin{array}{ll}
  f_{ib}^{P_{lr}}&=\alpha_lQ_l^{P_{lr}}+\beta_l+g_1(P_l)+\Re\big\{\sum_{k=1}^{\frac{N-1}{2}}\mu_{(l+k)_N}^*\mu_{(l-k)_N}^*\Phi_{l,k}\big\}\\
  f_{ib}^{P_{li}}&=\alpha_lQ_l^{P_{li}}+\beta_l+g_1(P_l)-\Re\big\{\sum_{k=1}^{\frac{N-1}{2}}\mu_{(l+k)_N}^*\mu_{(l-k)_N}^*\Phi_{l,k}\big\}\\
  f_{ib}^{\mu_{lr}}&=\alpha_lQ_l^{\mu_{lr}}+\Re\{j2\mu_{li}\sum_{k=1}^{\frac{N-1}{2}}\mu_{(l+k)_N}^*\mu_{(l-k)_N}^*\Phi_{l,k}\}\\
  &+\sum_{\substack{d=0\\d\neq l}}^{N-1}\Re\{\bar{P_d} T_{l,d}\}+\sum_{d=0}^{N-2}\sum_{k=1}^{\frac{N-1}{2}}\sum_{\substack{m=d+1\\m\neq (d+k)_N\\m\neq (d-k)_N}}^{N-1}\Re\{\Psi_{d,m,k}S_{d,m,k}^{\mu_{lr}}\}\\
  f_{ib}^{\mu_{li}}&=\alpha_lQ_l^{\mu_{li}}+\Re\{j2\mu_{lr}\sum_{k=1}^{\frac{N-1}{2}}\mu_{(l+k)_N}^*\mu_{(l-k)_N}^*\Phi_{l,k}\}\\
  &-\sum_{\substack{d=0\\d\neq l}}^{N-1}\Re\{j\bar{P_d} T_{l,d}\}+\sum_{d=0}^{N-2}\sum_{k=1}^{\frac{N-1}{2}}\sum_{\substack{m=d+1\\m\neq (d+k)_N\\m\neq (d-k)_N}}^{N-1}\Re\{\Psi_{d,m,k}S_{d,m,k}^{\mu_{li}}\}
\end{array}\right.,
\end{align}
where (we removed the details due to the lack of space)
\begin{align}\label{eqn_40}
g_1(P_l)&=\frac{d g(P_l)}{d P_l}=\sum_{\substack{m=0\\m\neq l}}^{N-1}\gamma_{m,l} P_m,\\
T_{d,l}&\triangleq \left\{\begin{array}{ll}
  \mu^{*}_{(2d-l)_N}\Phi_{l,(d-l)_N}~~\text{if}~~1\leq (d-l)_N \leq \frac{N-1}{2}\\
  \mu^{*}_{(2d-l)_N}\Phi_{l,(l-d)_N}~~\text{if}~~1\leq (l-d)_N \leq \frac{N-1}{2}
\end{array}\right.,\\\label{eqn_41}
Q_l^{P_{lr}}&=6P_{lr}+2P_{li},\\\label{eqn_42}
Q_l^{P_{li}}&=6P_{li}+2P_{lr},\\
Q_l^{\mu_{lr}}&=-8\mu_{lr}^3,\\
Q_l^{\mu_{li}}&=-8\mu_{li}^3,
\end{align}
and $S_{d,m,k}^{\mu_{lr}}$ and $S_{d,m,k}^{\mu_{li}}$ are defined as below
\begin{align}\nonumber
S_{d,m,k}^{\mu_{lr}}&=\delta_{l-d}\cdot\mu_{m}^{*}\mu_{(l-k)_N}^{*}\mu_{(m-k)_N}+\delta_{l-m}\cdot\mu_{d}\mu_{(d-k)_N}^{*}\mu_{(l-k)_N}\\
&+\delta_{l-(l-k)_N}\cdot\mu_{d}\mu_m^{*}\mu_{(m-k)_N}+\delta_{l-(m-k)_N}\cdot\mu_{d}\mu_m^{*}\mu_{(d-k)_N}^{*},\\\nonumber
S_{d,m,k}^{\mu_{li}}&=\delta_{l-d}\cdot j\mu_{m}^{*}\mu_{(l-k)_N}^{*}\mu_{(m-k)_N}-\delta_{l-m}\cdot j\mu_{d}\mu_{(d-k)_N}^{*}\mu_{(l-k)_N}\\
&-\delta_{l-(l-k)_N} \cdot j\mu_{d}\mu_m^{*}\mu_{(m-k)_N}+\delta_{l-(m-k)_N}\cdot j\mu_{d}\mu_m^{*}\mu_{(d-k)_N}^{*}.
\end{align}

It is verified that $\lambda_1$ can not be zero. Because otherwise the first two sets of equations in (\ref{eqn_12}) for the KKT conditions contradict the equalities. Therefore, the average power constraint is satisfied with equality due to (\ref{eqn_22}). For $\lambda_2=0$, it can be verified that we have the waterfilling solution.

\section{Proof of Lemma \ref{Lem_48}}\label{app_KKT_zeromean}

By noting that $\mu_l=0,~l=0,\ldots,N-1$ and writing the Lagrangian KKT conditions in (\ref{eqn_24}), we have
\begin{subequations}
\begin{align}
&\sum_{l=0}^{N-1}P_l\leq P_a,\\
&\sum_{l=0}^{N-1} f_{ib}(P_l,\bar{P_l},0,h_l,N)\geq P_d,\\
& P_{lr},P_{li}\geq 0,~l=0,...,N-1,\\
&\lambda_1\left(\sum_{l=0}^{N-1}P_l- P_a\right)=0,\\
&\lambda_2\left(P_d-\sum_{l=0}^{N-1} f_{ib}(P_l,\bar{P_l},0,h_l,N) \right)=0,\\
&\zeta_{lr}P_{lr}=\zeta_{li}P_{li}= 0,~l=0,...,N-1\\\label{eqn_43}
&\left\{\begin{array}{ll}
  \frac{-c_1a_l}{1+a_lP_{lr}}+\lambda_1-\lambda_2 f_{ib}^{P_{lr}}-\zeta_{lr}=0,~l=0,...,N-1,\\
  \frac{-c_1a_l}{1+a_lP_{li}}+\lambda_1-\lambda_2 f_{ib}^{P_{li}}-\zeta_{li}=0,~l=0,...,N-1,
\end{array}\right.
\end{align}
\end{subequations}
where
\begin{align}\nonumber
  f_{ib}^{P_{lr}}&=\alpha_lQ_l^{P_{lr}}+\beta_l+g_1(P_l)=\alpha_l(6P_{lr}+2P_{li})+\beta_l+g_1(P_l),\\\nonumber
  f_{ib}^{P_{li}}&=\alpha_lQ_l^{P_{li}}+\beta_l+g_1(P_l)=\alpha_l(6P_{li}+2P_{lr})+\beta_l+g_1(P_l).
\end{align}
with $g_1(P_l),~Q_l^{P_{lr}}$ and $Q_l^{P_{li}}$ defined in (\ref{eqn_40}), (\ref{eqn_41}) and (\ref{eqn_42}), respectively.

It is verified that $\lambda_1$ cannot be zero. Because otherwise the equations in (\ref{eqn_43}) contradict the equalities. Therefore, average power constraint is satisfied with equality. For $\lambda_2=0$ it can be verified that we have waterfilling solution. For $\lambda_2\neq 0$, note that $\zeta_{lr},~\zeta_{li},~l=0,\ldots,N-1$ act as slack variables, so they can be eliminated. We have
\begin{subequations}
\begin{align}
\sum_{l=0}^{N-1}P_l&= P_a,\\
\sum_{l=0}^{N-1} f_{ib}(P_l,\bar{P_l},0,h_l,N)&= P_d,\\
P_{lr},P_{li}&\geq 0,~l=0,...,N-1,\\\label{eqn_45}
P_{lr}\left(\lambda_1-G_l(\pmb{P}_{r}^{N},\pmb{P}_{i}^{N})\right)&=0,~l=0,...,N-1,\\\label{eqn_46}
P_{li}\left(\lambda_1-G_l(\pmb{P}_{i}^{N},\pmb{P}_{r}^{N})\right)&=0,~l=0,...,N-1,
\end{align}
\end{subequations}
where due to $\zeta_{lr},~\zeta_{li}\geq 0$ for any $l$, for the optimal solution, we have
\begin{align}
\lambda_1\geq \max\limits_{l} \{G_l(\pmb{P}_{r}^{N},\pmb{P}_{i}^{N}),G_l(\pmb{P}_{i}^{N},\pmb{P}_{r}^{N})\}.
\end{align}

\bibliographystyle{ieeetran}
\bibliography{ref}

\end{document}